\documentstyle[12pt]{article}

\def\be{\begin{equation}}
\def\ee{\end{equation}}
\def\bea{\begin{eqnarray}}
\def\eea{\end{eqnarray}}

\def\[{\lfloor{\hskip 0.35pt}\!\!\!\lceil}
\def\]{\rfloor{\hskip 0.35pt}\!\!\!\rceil}

\def\IR{\relax{\rm I\kern-.18em R}}
\def\IC{\relax{\rm I\kern-.18em C}}

\newcommand{\AmS}{{\protect\the\textfont2
  A\kern-.1667em\lower.5ex\hbox{M}\kern-.125emS}}

\catcode`@=11
\def\un#1{\relax\ifmmode\@@underline#1\else
        $\@@underline{\hbox{#1}}$\relax\fi}
\catcode`@=12

\def\fracm#1#2{\hbox{\large{${\frac{{#1}}{{#2}}}$}}}

\def\ad{{\kern0.5pt
                   \alpha \kern-5.05pt
\raise5.8pt\hbox{$\textstyle.$}\kern
0.5pt}}

\def\Dot#1{{\kern0.5pt
     {#1} \kern-5.05pt \raise5.8pt\hbox{$\textstyle.$}\kern
0.5pt}}

\def\a{\alpha}
\def\b{\beta}

\def\d{\delta}
\def\e{\epsilon}

\def\g{\gamma}

\def\l{\lambda}
\def\m{\mu}

\def\o{\omega}

\def\q{\theta}

\def\bo{{\raise.15ex\hbox{\large$\Box$}}}               
\def\pa{\partial}                                       
\def\TH{{\raise.2ex\hbox{$\displaystyle \bigodot$}\mskip-4.7mu \llap H
\;}}
\def\face{{\raise.2ex\hbox{$\displaystyle \bigodot$}\mskip-2.2mu \llap
{$\ddot
        \smile$}}}                                      

\def\Hat#1{\widehat{#1}}                        
\def\Bar#1{\overline{#1}}                       
\def\leftrightarrowfill{$\mathsurround=0pt \mathord\leftarrow \mkern-6mu
        \cleaders\hbox{$\mkern-2mu \mathord- \mkern-2mu$}\hfill
        \mkern-6mu \mathord\rightarrow$}
\def\dvec#1{\vbox{\ialign{##\crcr
        \leftrightarrowfill\crcr\noalign{\kern-1pt\nointerlineskip}
        $\hfil\displaystyle{#1}\hfil$\crcr}}}           

\def\fracm#1#2{\hbox{\large{${\frac{{#1}}{{#2}}}$}}}
\def\frac#1#2{{\textstyle{#1\over\vphantom2\smash{\raise.20ex
        \hbox{$\scriptstyle{#2}$}}}}}                   
\def\sfrac#1#2{{\vphantom1\smash{\lower.5ex\hbox{\small$#1$}}\over
        \vphantom1\smash{\raise.4ex\hbox{\small$#2$}}}} 
\def\bfrac#1#2{{\vphantom1\smash{\lower.5ex\hbox{$#1$}}\over
        \vphantom1\smash{\raise.3ex\hbox{$#2$}}}}       
\def\afrac#1#2{{\vphantom1\smash{\lower.5ex\hbox{$#1$}}\over#2}}    

\newskip\humongous \humongous=0pt plus 1000pt minus 1000pt
\def\caja{\mathsurround=0pt}
\def\eqalign#1{\,\vcenter{\openup2\jot \caja
        \ialign{\strut \hfil$\displaystyle{##}$&$
        \displaystyle{{}##}$\hfil\crcr#1\crcr}}\,}
\newif\ifdtup

  \def\pp{{\mathchoice
          {
              \kern 1pt%
              \raise 1pt
              \vbox{\hrule width5pt height0.4pt depth0pt
                    \kern -2pt
                    \hbox{\kern 2.3pt
                          \vrule width0.4pt height6pt depth0pt
                          }
                    \kern -2pt
                    \hrule width5pt height0.4pt depth0pt}%
                    \kern 1pt
           }
            {
              \kern 1pt%
              \raise 1pt
              \vbox{\hrule width4.3pt height0.4pt depth0pt
                    \kern -1.8pt
                    \hbox{\kern 1.95pt
                          \vrule width0.4pt height5.4pt depth0pt
                          }
                    \kern -1.8pt
                    \hrule width4.3pt height0.4pt depth0pt}%
                    \kern 1pt
            }
            {
              \kern 0.5pt%
              \raise 1pt
              \vbox{\hrule width4.0pt height0.3pt depth0pt
                    \kern -1.9pt  
                    \hbox{\kern 1.85pt
                          \vrule width0.3pt height5.7pt depth0pt
                          }
                    \kern -1.9pt
                    \hrule width4.0pt height0.3pt depth0pt}%
                    \kern 0.5pt
            }
            {
              \kern 0.5pt%
              \raise 1pt
              \vbox{\hrule width3.6pt height0.3pt depth0pt
                    \kern -1.5pt
                    \hbox{\kern 1.65pt
                          \vrule width0.3pt height4.5pt depth0pt
                          }
                    \kern -1.5pt
                    \hrule width3.6pt height0.3pt depth0pt}%
                    \kern 0.5pt
            }
        }}

  \def\mm{{\mathchoice
                       {
                             \kern 1pt
               \raise 1pt    \vbox{\hrule width5pt height0.4pt depth0pt
                                  \kern 2pt
                                  \hrule width5pt height0.4pt depth0pt}
                             \kern 1pt}
                       {
                            \kern 1pt
               \raise 1pt \vbox{\hrule width4.3pt height0.4pt depth0pt
                                  \kern 1.8pt
                                  \hrule width4.3pt height0.4pt depth0pt}
                             \kern 1pt}
                       {
                            \kern 0.5pt
               \raise 1pt
                            \vbox{\hrule width4.0pt height0.3pt depth0pt
                                  \kern 1.9pt
                                  \hrule width4.0pt height0.3pt depth0pt}
                            \kern 1pt}
                       {
                           \kern 0.5pt
             \raise 1pt  \vbox{\hrule width3.6pt height0.3pt depth0pt
                                  \kern 1.5pt
                                  \hrule width3.6pt height0.3pt depth0pt}
                           \kern 0.5pt}
                       }}

\def\pd{{\kern0.5pt
                   + \kern-5.05pt \raise5.8pt\hbox{$\textstyle.$}\kern
0.5pt}}

\def\pmd{{\kern0.5pt
                  \pm \kern-5.05pt \raise6.3pt\hbox{$\textstyle.$}\kern1.5pt}}

\def\md{{\mathchoice
   {
      {{\kern 1pt - \kern-6.2pt \raise5pt\hbox{$\textstyle.$}\kern 1pt}}}
    {
      {{\kern 1pt - \kern-6.2pt \raise5pt\hbox{$\textstyle.$}\kern 1pt}}}
    {
      {\kern0.5pt - \kern-5.05pt \raise3.4pt\hbox{$\textstyle.$}\kern0.5pt}}
    {
      {\kern0.5pt - \kern-5.05pt \raise3.4pt\hbox{$\textstyle.$}\kern0.5pt}}}}

\def\ad{{\dot{\alpha}}}

\def\pp{{\mathchoice
          {
              \kern 1pt%
              \raise 1pt
              \vbox{\hrule width5pt height0.4pt depth0pt
                    \kern -2pt
                    \hbox{\kern 2.3pt
                          \vrule width0.4pt height6pt depth0pt
                          }
                    \kern -2pt
                    \hrule width5pt height0.4pt depth0pt}%
                    \kern 1pt
           }
            {
              \kern 1pt%
              \raise 1pt
              \vbox{\hrule width4.3pt height0.4pt depth0pt
                    \kern -1.8pt
                    \hbox{\kern 1.95pt
                          \vrule width0.4pt height5.4pt depth0pt
                          }
                    \kern -1.8pt
                    \hrule width4.3pt height0.4pt depth0pt}%
                    \kern 1pt
            }
            {
              \kern 0.5pt%
              \raise 1pt
              \vbox{\hrule width4.0pt height0.3pt depth0pt
                    \kern -1.9pt  
                    \hbox{\kern 1.85pt
                          \vrule width0.3pt height5.7pt depth0pt
                          }
                    \kern -1.9pt
                    \hrule width4.0pt height0.3pt depth0pt}%
                    \kern 0.5pt
            }
            {
              \kern 0.5pt%
              \raise 1pt
              \vbox{\hrule width3.6pt height0.3pt depth0pt
                    \kern -1.5pt
                    \hbox{\kern 1.65pt
                          \vrule width0.3pt height4.5pt depth0pt
                          }
                    \kern -1.5pt
                    \hrule width3.6pt height0.3pt depth0pt}%
                    \kern 0.5pt
            }
        }}

  \def\mm{{\mathchoice
                       {
                             \kern 1pt
               \raise 1pt    \vbox{\hrule width5pt height0.4pt depth0pt
                                  \kern 2pt
                                  \hrule width5pt height0.4pt depth0pt}
                             \kern 1pt}
                       {
                            \kern 1pt
               \raise 1pt \vbox{\hrule width4.3pt height0.4pt depth0pt
                                  \kern 1.8pt
                                  \hrule width4.3pt height0.4pt depth0pt}
                             \kern 1pt}
                       {
                            \kern 0.5pt
               \raise 1pt
                            \vbox{\hrule width4.0pt height0.3pt depth0pt
                                  \kern 1.9pt
                                  \hrule width4.0pt height0.3pt depth0pt}
                            \kern 1pt}
                       {
                           \kern 0.5pt
             \raise 1pt  \vbox{\hrule width3.6pt height0.3pt depth0pt
                                  \kern 1.5pt
                                  \hrule width3.6pt height0.3pt depth0pt}
                           \kern 0.5pt}
                       }}

\def\pd{{\kern0.5pt
                   + \kern-5.05pt \raise5.8pt\hbox{$\textstyle.$}\kern
0.5pt}}

\def\pmd{{\kern0.5pt
                  \pm \kern-5.05pt \raise6.3pt\hbox{$\textstyle.$}\kern1.5pt}}

\def\md{{\mathchoice
   {
      {{\kern 1pt - \kern-6.2pt \raise5pt\hbox{$\textstyle.$}\kern 1pt}}}
    {
      {{\kern 1pt - \kern-6.2pt \raise5pt\hbox{$\textstyle.$}\kern 1pt}}}
    {
      {\kern0.5pt - \kern-5.05pt \raise3.4pt\hbox{$\textstyle.$}\kern0.5pt}}
    {
      {\kern0.5pt - \kern-5.05pt \raise3.4pt\hbox{$\textstyle.$}\kern0.5pt}}}}

\def\dslash{\not{\hbox{\kern-2pt $\partial$}}}
\def\Dslash{\not{\hbox{\kern-4pt $D$}}}
\def\pslash{\not{\hbox{\kern-2.3pt $p$}}}
 \newtoks\slashfraction
 \slashfraction={.13}
 \def\slash#1{\setbox0\hbox{$ #1 $}
 \setbox0\hbox to \the\slashfraction\wd0{\hss \box0}/\box0 }
 
\font\ro=cmsy10                          
\def\kcr{{\hbox{\ro \char'170}}}                
\def\ktl{{\hbox{\ro \char'170}}}        
\def\ktr{{\hbox{\ro \char'170}}}        
\def\kbl{{\hbox{\ro \char'170}}}        
\def\kbr{{\hbox{\ro \char'170}}}        

\def\plpl{\raise-2pt\hbox{$\raise3pt\hbox{$_+$}\hskip-6.67pt\raise0.0pt
\hbox{$^+$}\hskip 0.01pt$}}
\def\mimi{\raise-2pt\hbox{$\raise3pt\hbox{$_-$}\hskip-6.67pt\raise0.0pt
\hbox{$^-$}\hskip 0.01pt$}} 

\def\bo{{\raise.15ex\hbox{\large$\Box$}}}               
\def\pa{\partial}                                       
\def\TH{{\raise.2ex\hbox{$\displaystyle \bigodot$}\mskip-4.7mu \llap H \;}}
\def\face{{\raise.2ex\hbox{$\displaystyle \bigodot$}\mskip-2.2mu \llap {$\ddot
        \smile$}}}                                      

\def\Hat#1{\widehat{#1}}                        
\def\Bar#1{\overline{#1}}                       
\def\leftrightarrowfill{$\mathsurround=0pt \mathord\leftarrow \mkern-6mu
        \cleaders\hbox{$\mkern-2mu \mathord- \mkern-2mu$}\hfill
        \mkern-6mu \mathord\rightarrow$}
\def\dvec#1{\vbox{\ialign{##\crcr
        \leftrightarrowfill\crcr\noalign{\kern-1pt\nointerlineskip}
        $\hfil\displaystyle{#1}\hfil$\crcr}}}           

\def\fracm#1#2{\hbox{\large{${\frac{{#1}}{{#2}}}$}}}
\def\frac#1#2{{\textstyle{#1\over\vphantom2\smash{\raise.20ex
        \hbox{$\scriptstyle{#2}$}}}}}                   
\def\sfrac#1#2{{\vphantom1\smash{\lower.5ex\hbox{\small$#1$}}\over
        \vphantom1\smash{\raise.4ex\hbox{\small$#2$}}}} 
\def\bfrac#1#2{{\vphantom1\smash{\lower.5ex\hbox{$#1$}}\over
        \vphantom1\smash{\raise.3ex\hbox{$#2$}}}}       
\def\afrac#1#2{{\vphantom1\smash{\lower.5ex\hbox{$#1$}}\over#2}}    

\parskip=\medskipamount                 
\thicklines                         

\def\oldheadpic{                                
        \setlength{\unitlength}{.4mm}
        \thinlines
        \par
        \begin{picture}(349,16)
        \put(325,16){\line(1,0){4}}
        \put(330,16){\line(1,0){4}}
        \put(340,16){\line(1,0){4}}
        \put(335,0){\line(1,0){4}}
        \put(340,0){\line(1,0){4}}
        \put(345,0){\line(1,0){4}}
        \put(329,0){\line(0,1){16}}
        \put(330,0){\line(0,1){16}}
        \put(339,0){\line(0,1){16}}
        \put(340,0){\line(0,1){16}}
        \put(344,0){\line(0,1){16}}
        \put(345,0){\line(0,1){16}}
        \put(329,16){\oval(8,32)[bl]}
        \put(330,16){\oval(8,32)[br]}
        \put(339,0){\oval(8,32)[tl]}
        \put(345,0){\oval(8,32)[tr]}
        \end{picture}
        \par
        \thicklines
        \vskip.2in}
\def\oldtitle#1#2#3#4{\oldheadpic\begin{center}\vglue.5in{\large\bf #1}\\[.6in]
        {#2}\\[.1in] {\it Department of Physics and Astronomy}\\
        {\it University of Maryland, College Park, MD 20742}\\[.6in]
        Physics Publication \#{#3}\\ {#4}\\[1.5in] {\bf ABSTRACT}\\[.1in]
        \end{center} \begin{quotation}}                 
\def\oldTitle#1#2#3#4#5#6#7{\oldheadpic\begin{center} \vglue .4in
        {\large\bf #1}\\[.4in]
        {#2}\\[.1in] {\it Department of Physics and Astronomy}\\
        {\it University of Maryland, College Park, MD 20742}\\[.1in]
        {#3}\\[.1in] {\it {#4}}\\ {\it {#5}}\\[.4in]
        Physics Publication \#{#6}\\ {#7}\\[.5in] {\bf ABSTRACT}\\[.1in]
        \end{center} \begin{quotation}}                 
\def\border{                                            
        \setlength{\unitlength}{1mm}
        \newcount\xco
        \newcount\yco
        \xco=-21
        \yco=12
        \begin{picture}(140,0)
        \put(\xco,\yco){$\ktl$}
        \advance\yco by-1
        {\loop
        \put(\xco,\yco){$\kcr$}
        \advance\yco by-2
        \ifnum\yco>-240
        \repeat
        \put(\xco,\yco){$\kbl$}}
        \xco=158
        \yco=12
        \put(\xco,\yco){$\ktr$}
        \advance\yco by-1
        {\loop
        \put(\xco,\yco){$\kcr$}
        \advance\yco by-2
        \ifnum\yco>-240
        \repeat
        \put(\xco,\yco){$\kbr$}}
        \put(-20,13){\tiny University of Maryland Elementary Particle
Physics University of Maryland Elementary Particle Physics University of
Maryland Elementary Particle Physics}
        \put(-20,-241.5){\tiny University of Maryland Elementary
Particle Physics University of Maryland Elementary Particle Physics
University of Maryland Elementary Particle Physics}
        \end{picture}
        \par\vskip-8mm}
\def\bordero{                                           
        \setlength{\unitlength}{1mm}
        \newcount\xco
        \newcount\yco
        \xco=-31
        \yco=12
        \begin{picture}(140,0)
        \put(\xco,\yco){$\ktl$}
        \advance\yco by-1
        {\loop
        \put(\xco,\yco){$\kclr}
        \advance\yco by-2
        \ifnum\yco>-240
        \repeat
        \put(\xco,\yco){$\kbl$}}
        \xco=151
        \yco=12
        \put(\xco,\yco){$\ktr$}
        \advance\yco by-1
        {\loop
        \put(\xco,\yco){$\kcr$}
        \advance\yco by-2
        \ifnum\yco>-240
        \repeat
        \put(\xco,\yco){$\kbr$}}
        \put(-20,12){\ooo bacdefghidfghghdhededbihdgdfdfhhdheidhdhebaaahjhhdahba

hgdedge
   hgfdiehhgdigicba}
        \put(-20,-241.5){\ooo ababaighefdbfghgeahgdfgafagihdidihiidhiagfedhadbfd

ecdcdfa
   gdcbhaddhbgfchbgfdacfediacbabab}
        \end{picture}
        \par\vskip-8mm}
\def\headpic{                                           
        \indent
        \setlength{\unitlength}{.4mm}
        \thinlines
        \par
        \begin{picture}(29,16)
        \put(165,16){\line(1,0){4}}
        \put(170,16){\line(1,0){4}}
        \put(180,16){\line(1,0){4}}
        \put(175,0){\line(1,0){4}}
        \put(180,0){\line(1,0){4}}
        \put(185,0){\line(1,0){4}}
        \put(169,0){\line(0,1){16}}
        \put(170,0){\line(0,1){16}}
        \put(179,0){\line(0,1){16}}
        \put(180,0){\line(0,1){16}}
        \put(184,0){\line(0,1){16}}
        \put(185,0){\line(0,1){16}}
        \put(169,16){\oval(8,32)[bl]}
        \put(170,16){\oval(8,32)[br]}
        \put(179,0){\oval(8,32)[tl]}
        \put(185,0){\oval(8,32)[tr]}
        \end{picture}
        \par\vskip-6.5mm
        \thicklines}
\def\title#1#2#3#4{\border\headpic {\hbox to\hsize{#4 \hfill UMDEPP #3}}\par
        \begin{center} \vglue .5in {\large\bf #1}\\[.6in]
        {#2}\\[.1in] {\it Department of Physics and Astronomy}\\
        {\it University of Maryland, College Park, MD 20742}\\[1.5in]
        {\bf ABSTRACT}\\[.1in] \end{center} \begin{quotation}}  
\def\Title#1#2#3#4#5#6#7{\border\headpic
        {\hbox to\hsize{#7 \hfill UMDEPP #6}}\par
        \begin{center} \vglue .4in {\large\bf #1}\\[.4in]
        {#2}\\[.1in] {\it Department of Physics and Astronomy}\\
        {\it University of Maryland, College Park, MD 20742}\\[.1in]
        {#3}\\[.1in] {\it {#4}}\\ {\it {#5}}\\[.5in] {\bf ABSTRACT}\\[.1in]
        \end{center} \begin{quotation}}                 
\def\endtitle{\end{quotation}\newpage}                  

\def\qd{{\kern0.5pt q \kern-5.05pt \raise5.8pt\hbox{$\textstyle.$}\kern
0.5pt}}

\begin{document}

\def\gfrac#1#2{\frac {\scriptstyle{#1}}
        {\mbox{\raisebox{-.6ex}{$\scriptstyle{#2}$}}}}
\def\gg{{\hbox{\sc g}}}
\border\headpic {\hbox to\hsize{June 2001 \hfill
{UMDEPP 01-058}}}
\par
\setlength{\oddsidemargin}{0.3in}
\setlength{\evensidemargin}{-0.3in}
\begin{center}
\vglue .1in
{\Large\bf Superconformal Symmetry in \\
11D Superspace and the \\ \vglue .15in
M-Theory Effective Action\footnote{
Supported in part by National Science Foundation Grants 
PHY-98-02551}  } \\[.25in]
S.\ James Gates, Jr.\footnote{gatess@wam.umd.edu}
\\[0.06in]
{\it Department of Physics\\ University of Maryland\\ 
College Park, MD 20742-4111 USA}
\\[1.5in]

{\bf ABSTRACT}\\[.01in]
\end{center}
~~~~We establish a theorem about non-trivial 11D supergravity fluctuations 
that are conformally related to flat superspace geometry.  Under the
assumption that a theory of conformal 11D supergravity exists, similar
in form to that of previously constructed theories in lower dimensions, 
this theorem demands the appearance of non-vanishing dimension 1/2 
torsion tensors in order to accommodate a non-trivial 11D conformal
compensator and thus M-theory corrections that break superconformal
symmetry.   At the complete non-linear level, a presentation of a
conventional minimal superspace  realization of Weyl symmetry in eleven
dimensional superspace is also described.  All of our results taken 
together imply that there exists some realization of conformal symmetry 
relevant for the M-theory effective action.  We thus led to conjecture 
this is also true for the full and complete M-theory.
\begin{quotation}
{}  

${~~~}$ \newline
PACS: 03.70.+k, 11.30.Rd, 04.65.+e    

Keywords: Gauge theories, Supersymmetry, Supergravity.
\endtitle

\section {Introduction}  

~~~~In a recent note \cite{GN01}, we have resumed studies of a class 
of problems, the on-shell superspace and perturbative description of 
higher derivative supergravity, that has been one of several foci of 
fascination for us since we inaugurated such investigations
\cite{G1,G2,G3,G4,G5} in the middle eighties.  Our proposed method 
uses on-shell superspace to describe, in a perturbative manner, 
supergravity actions containing higher-order curvature terms.  As 
first stated in the initial efforts, in order for such constructions
to make sense there must be a dimensional parameter with respect to 
which such a perturbative description is made.  One might think that 
such a parameter is present in {\it {all}} ordinary supergravity systems.  
After all, the Newton constant is always present in such theories.  In 
fact this is not sufficient.  When supergravity theories are written 
in superspace formulations, it is always possible to perform certain 
implicit ``re-scalings'' of component fields in such a way that 
Newton's constant is effectively absent.  One requires a second 
dimensional constant and this is supplied in superstring theory by 
the string tension.  Similarly, in discussions of the low-energy 
M-theory effective action there is also such a constant that we
refer to as $\ell_{11}$.  We can choose this parameter to possess
the units of length or that of an inverse mass.  As proposed
\cite{G1,G2,G3,G4,G5} in our inaugural papers on this topic, the
superspace torsion, curvature and field strength can be `deformed'
perturbatively as power series expansions in terms of such a dimensional
parameter. Although our arguments were made with regard to 
superstring/heterotic string theories, the same reasoning applies to 
superstring/M-theory.

Many years ago, we made a conjecture \cite{G80} about some of the 
structure that is required to describe an eleven dimensional 
supergravity theory whose equations of motion is different from those
derived from the standard Cremmer-Julia theory.  In fact in 1980, it 
was pointed out that such modification would in all likelihood require 
the existence of a dimension 1/2 and spin-1/2 multiplet  of currents 
(which we now refer to as the ${\cal J}$-tensor).  At that time an 
equation was given for how this multiplet of currents would begin to 
modify the supergeometry of eleven dimensional supergravity.  In a 
1996 work \cite{G6} we attempted to extend this investigation.   However,
due to missing terms\footnote{Interestingly, these terms were already
present in the 1980 work.} first noted by  Howe \cite{HOW}, this was 
not a carried out convincingly.  In fact,  Howe {\it {even}} argued that 
our proposed dimension 1/2 and spin-1/2 multiplet could not play a  
role in 11D supergravity/M-theory.  This was formalized in a result
sometimes called ``Howe's Theorem.'' Most recently \cite{GN01}, we 
have {\em {emphasized}} that Howe's Theorem  cannot be valid unless
supergravity/M-theory dynamics admits a scale-invariance.  Since the
proposed lowest order M-theory correction violates this condition, 
it is fully our expectation that the dimension 1/2 and spin-1/2 
multiplet of currents must play a critical role.

In the mean time, research has been undertaken based upon ``Howe's 
Theorem'' and investigating what modifications are allowed within its 
context.   This has led to the appearance of additional multiplets of 
currents known as $X$-tensors to be introduced so as to accommodate 
the higher derivative M-theory corrections.  The appearance of the
$X$-tensors  does not contradict the appearance of the spin-1/2 ${\cal
J}$-tensor.  In fact,  in both our 1980 and 1996 discussions of the 
modified 11D supergeometry we made explicit reference to our expectation 
that the spin-1/2 ${\cal J}$-tensor would most likely be {\it {only}} 
part of an off-shell theory i.\ e.\ the ${\cal J}$-tensor must be
accompanied by other tensors to provide a superspace description of
the M-theory corrections.

One of the most powerful argument that reveals why the $X$-tensors
alone are {\em {insufficient}} to describe an off-shell Poincar\' e 
supergravity theory is because they are actually superconformal
field strengths\footnote{This fact seems not to have been noticed 
until the recent work in \cite{GN01}.}, similar to but distinct 
from, the superspace Weyl multiplet superfield.  We should also note 
that the appearance of the superconformal $X$-tensors was also 
foreshadowed in a much overlooked study of conformal symmetry in 4D, 
$N > 4$ superspace geometry \cite{GG}. In this work there appears 
the statement, ``The present analysis shows that conformal symmetry 
may still be relevant for $N > 4$ supergravity.  However, it 
requires at least a {\em {second}} conformal field strength...''   
Thus, by this work, the realization that a superconformal superspace 
field strength not related to the Weyl tensor has appeared in the
literature for some time.  Although our comments were directed to 
4D, $N > 4$ superspace geometry, dimensional oxidation implies that 
this must be true for 11D superspace geometry also.  So the significance 
of our observation has largely been ignored and has led to much 
confusion on this subject.

\section {Conformal Class of Bosonic Spaces}

~~~~A special class of infinitesimal fluctuations are those that
are conformally related to the flat metric of eleven dimensional
superspace.  We begin our study of eleven dimensional supergeometry
with this class of theories.  However, before we do so, we wish
to review this class of structures within the non-supersymmetric
case followed by considerations in a well-known supersymmetric 
context.

In a purely bosonic space, the class of graviton fluctuations that 
are conformally related to flat space are defined by
\be
\nabla_a ~=~ \pa_a ~+~ \psi \pa_a ~+~ k_0 \, (\pa_b \psi) {\cal M
}_a {}^b ~~~,
\label{eq:Clss-5} \ee
where the infinitesimal scalar field $\psi$ (an arbitrary function 
of the spacetime coordinates) is known as the ``scale factor,'' ``scale
compensator,'' or ``conformal compensator.''  Torsion and  curvature
tensors can be defined by calculating the commutator algebra of this
derivative.
\be \eqalign{
[\, \nabla_a ~	,~ \nabla_b \, ] \, &=~ t_{a \, b}{}^c \, \nabla_c 
~+~ \fracm 12 \, r_{a \, b ~ c}{}^d \, {\cal M}_d {}^c  ~~~, \cr
t_{a \, b}{}^c \, &=~  (\, 1 ~+~ k_0  \,) \,(\pa_{[a} \psi) \,
\d_{b]}{}^c  ~~~,~~~ \cr 
r_{a \, b ~}{}^{d e}  \, &=~  k_0 \, [~ (\pa^{d} \pa_{[a} \psi) \,
\d_{b]}{}^e  ~-~  (\pa^{e} \pa_{[a} \psi) \, \d_{b]}{}^d ~ ]  ~~~.
}\label{eq:Clss-4} \ee
A Riemannian geometry has no torsion tensor and this is  
accomplished by demanding,
\be
 t_{a \, b}{}^c ~=~ 0 ~~\to~~~ k_0 ~=~ - 1 ~~~,
\label{eq:Clss-4X} \ee
thus determining the value of the otherwise undetermined parameter
$k_0$ above.  As the vanishing torsion determines the Christoffel
connection, we see that fixing $k_0 = -1$, is equivalent to choosing
the Christoffel connection in (\ref{eq:Clss-5}).  Geometries that 
are conformally related to flat geometry are also compatible with 
a Riemannian geometry possessing completely vanishing torsion. Stated 
another way, the vanishing of the torsion places no dynamical
restriction on the compensator $\psi$.

For finite values of the conformal compensator, we can integrate
(\ref{eq:Clss-5}) to obtain
\be
\nabla_a ~=~\psi \, [ ~ \pa_a ~-~ \, (\pa_b \, ln \psi) {\cal M
}_a {}^b ~] ~~~.
\label{eq:Clss-5F} \ee

\section {Minimal Conformal Class of 4D, $N$ = 1 \\Superspace}

~~~~We can continue this line of deliberation by looking at the case
of 4D, $N$ = 1 superspace.  The analog of (\ref{eq:Clss-5})
takes the form
\be \eqalign{
\nabla_{\a} &=~  D_{\a} ~+~ \fracm 12 \, \Psi \, D_{\a}
~+~ \ell_0 (D_{\b} \Psi) {\cal M}_{\a}{}^{\b}  ~~~, \cr
\nabla_{\un a} &=~  \pa_{\un a} ~+~ \fracm 12 \,(\, \Psi 
\,+\, {\Bar {\Psi}} \,) \, \pa_{\un a} ~+~ i\, \ell_1 
\,  (D_{\a} {\Bar \Psi} ) \,  {\Bar D}{}_{\Dot \a} \,+\, i\, 
{\Bar \ell}{}_1 \, ({\Bar D}{}_{\Dot \a} \Psi ) 
\, D_{\a} \cr
&{~~~}~+~ i {\ell}{}_2\,  \, ([\, D_{\g} ~,~ {\Bar D}{}_{\Dot \a} 
\,]\, \Psi ) \, {\cal M}{}_{\a}{}^{\g} ~-~i {\Bar \ell}{}_2 ~\, 
([\, D_{\a} ~,~ {\Bar D}{}_{\Dot \g} \,] \, {\Bar \Psi}) \, {\Bar 
{\cal M}}{}_{\Dot \a}{}^{\Dot \g} \cr 
&{~~~}~+~ {\ell}{}_3 \, (\pa_{\g \Dot \a} \Psi) \, {\cal M}_{\a 
}{}^{\g} ~+~ {\Bar \ell}{}_3 \, (\pa_{\a \Dot \g} {\Bar \Psi}) 
\, {\Bar {\cal M}}{ }_{\Dot \a}{}^{\Dot \g}   ~~~, 
}  \label{eq:Clss-1XY}  \ee
In this expression, the constants $\ell_0$, $\ell_1$, $\ell_2$ and 
$\ell_3$ are the supersymmetric analogs of $k_0$ in (\ref{eq:Clss-5})
and $\Psi$ is a {\em {complex}} scalar and an infinitesimal superfield.
It is an exercise to compute all the  dimension 1/2 torsion tensors 
associated with (\ref{eq:Clss-1XY}) and we find
\be  \eqalign{
T_{\a \, \b}{}^{\g} \, &=~ \fracm 12 \, (\, \ell_0 \,+\, 1 
\,) \, (D_{(\a} \Psi) \, \d_{\b )}{}^{\g} ~~~,  \cr  
T_{\a \, \Dot \b}{}^{\Dot \g} \, &=~ (\, \ell_1 \,+\, 
\fracm 12 \,)\, (D_{\a} {\Bar \Psi}) \, \d_{\Dot \b}{}^{\Dot \g} 
~~~, \cr  
T_{\a \, {\un b}}{}^{\un c} \, &=~ \{ ~ \fracm 12 [ \, D_{\a} (\, 
(\, 1\,-\, \ell_0 \,) \Psi ~+~ {\Bar \Psi}  \, ) \,] \, \d_{\b}{
}^{\g}  \cr 
&{~~~~~}~+~ [ \, D_{\b}( \, \ell_0  \Psi\,+\, \ell_1 {\Bar \Psi}
\, )\,] \, \d_{\a}{}^{\g}  ~\} \, \d_{\Dot \b}{}^{\Dot \g}  ~~~.
}\label{eq:Clss-3X} \ee 
The constraints of 4D, $N$ = 1 supergravity act much as their
analog in (\ref{eq:Clss-4X}).  The condition
\be
\nabla_{\un a} ~=~ - i \, [\, \nabla_{\a} ~,~ \nabla_{\Dot \a} \, \} ~~~,
\label{eq:Clss-5X} \ee
determines the constants $\ell_1$, $\ell_2$ and $\ell_3$ in terms of
$\ell_0$.  This is equivalent to providing a definition of E${}_a$ and 
$\o_{\a ~ \un b \, \un c}$. The condition
\be
T_{\a \, \b}{}^{\g} ~=~ 0 ~~~,
\label{eq:Clss-6X} \ee
determines the constant  $\ell_0$ and this is equivalent to providing a
definition of $\o_{\a ~ \b \, \g}$.  The form of the fluctuations we are
considering also require that $\o_{\a ~ {\Dot \b} \, {\Dot \g}} = 0$.

Demanding that the first two dimension 1/2 torsions should vanish leads 
to $\ell_0 = - 1$ and $\ell_1 = - \fracm 12$.  In principle in order for 
the {\em {last}} dimension 1/2 torsion to vanish, we are led to {\em {two}}
{\em {distinct}} conditions
\be 
D_{\a} [\, (\, 1\,-\, \ell_0 \,) \Psi ~+~ {\Bar \Psi}  \, ] ~=~ 0 ~~~,
~~~ D_{\b}[ \, \ell_0  \Psi\,+\, \ell_1 {\Bar \Psi} \,] ~=~ 0 ~~~,
\label{eq:Clss-4X0} \ee
except for $\ell_0 = - 1$ and $\ell_1 = - \fracm 12$ these are proportional 
to the self-same equation
\be
D_{\a} (\, 2 \Psi ~+~ {\Bar {\Psi}} \,) ~=~ 0 ~~~.
\label{eq:Clss-4X1} \ee
A solution of this equation introduces the well-known ``chiral
compensator'' superfield, $\varphi$, of 4D, $N$ = 1 supergravity
\cite{ChrLcomp}.
\be
\Psi  ~=~  (~ 2 {\Bar {\varphi}} ~-~ \varphi ~ )
~~~.
\label{eq:Clss-4XX} \ee
We see that 4D, $N$ = 1 supergravity fluctuations, conformally related 
to flat superspace, do {\em {not}} require non-vanishing dimension 1/2 
torsions {\it{if}} {\it{and}} {\it{if}} those fluctuations are described 
by a chiral compensator. The vanishing of the dimension 1/2 torsion places
no dynamical restriction on the chiral compensator.  The results of
(\ref{eq:Clss-1XY}) - (\ref{eq:Clss-4XX}) for the special values $\ell_0 
= - 1$ and $\ell_1 = - \fracm 12$ correspond precisely to the infinitesimal
limit  of the minimal 4D, $N$ = 1 prepotential formulation of supergravity 
with the added restriction that we are working in a gauge where the
conformal prepotential, $U^{\un m}$, has also been set to zero.  The
property that  the full supergravity solution allows us to consistently
separate the  fluctuations in $\nabla_{\un A}$ involving $U^{\un m}$ from
those involving $\varphi$ may be referred to as ``separability.''
Separability is equivalent to the statement that a Poincar\' e supergravity
multiplet can be thought of as the combination of a Weyl supergravity
multiplet and a compensator multiplet.  The compensator multiplet contains
all of the Goldstone fields required to break the superconformal symmetry
group to the super Poincar\' e group.

\section {The Conformal Class of Flat 11D Superspace}

~~~~We now wish to prove a theorem about the eleven dimensional
supersymmetric case.  We call this the ``11D Torsion--Conformal Compensator
Theorem'' (11D T--C${}^2$ Theorem).  Below we will show that 11D
supergravity has  a behavior that is drastically different from either the
purely bosonic  theory or 4D, $N$ = 1 supergravity discussed above.  The
formal statement  of the 11D T--C${}^2$ Theorem is given below.

${~~~~}$ {\it {If 11D supergravity is completely separable into a 
Weyl supermulti- \newline ${~~\,~~~~~}$ plet and a compensator 
supermultiplet, then the 11D Poincar\' e super- \newline
${~~\,~~~~~}$ geometry {\underline {must}}
possess {\underline {non-vanishing}} dimension 1/2 
torsion super- \newline
${~~\,~~~~~}$ tensors.}}
\vglue .06in  \noindent We now prove this by explicit construction.  
The condition of complete separability means that we can set to zero 
the superfield that contains the Weyl multiplet without affecting the 
dynamics of the conformal  compensator.   This condition is true in 
{\em {all}} superfield  supergravity theories {\it {presently}} known.

For the class of eleven dimensional superspaces, we note that the analog 
of (\ref{eq:Clss-5}) and (\ref{eq:Clss-1XY}) takes the form
\be \eqalign{
\nabla_{\a} &=~  D_{\a} ~+~ \fracm 12 \, \Psi \, D_{\a}
~+~ \ell_0 (D_{\b} \Psi) (\g^{d e})_{\a}{}^{\b} 
{\cal M}_{ d e}  ~~~, \cr
\nabla_{a} &=~  \pa_{a} ~+~ \Psi \, \pa_{a} ~+~ i\, \ell_1 \,
(\g_a)^{\a \, \b} \, (D_{\a} \Psi ) \,  D_{\b} ~+~ \ell_2 \, 
(\pa_c \Psi) \, {\cal M}_{a}{}^c  \cr
&{~~~}~+~i \ell_3 \, (\g_{a}^{~d e} )^{\a \b} \, (D_{\a}
D_{\b} \Psi)  \, {\cal M}_{d e}  ~~~,
}  \label{eq:Clss-1}  \ee
where $\Psi$ in (\ref{eq:Clss-1})\footnote{Although we use the same
symbol here, this superfield should not be confused with the \newline
${~~~~~}$ one that appears in (\ref{eq:Clss-1XY}).}  is a {\em {real}}
scalar and an infinitesimal superfield and the $\ell_i$'s are a set of
constants (essential like those of the 4D, $N$ = 1 theory).  Of course 
the form of E${}_{\a}$ is given by
\be
{\rm E}{}_{\a} ~=~ D_{\a} ~+~ \fracm 12 \Psi D_{\a} ~~~,
\ee
which is appropriate for a supergeometry that is related
by a scale transformation to a ``flat'' eleven dimensional
supergeometry.  The constants $\ell_2$ and $\ell_3$ can be
eliminated by imposing the analog of (\ref{eq:Clss-5X}) to the
11D theory.  This leaves only $\ell_0$ and $\ell_1$ to be fixed.

Computing the dimension 1/2 torsion tensors associated with
(\ref{eq:Clss-1}), we find
\be  \eqalign{ 
T_{\a \, b}{}^c \, &=~ (\, 1 ~+~ 2 \ell_0\,) \, (D_{\a} \Psi) 
\, \d_b {}^c ~-~ (\, \ell_1 ~+~ 2 \ell_0 \,) (\g^c \, \g_b)_{
\a}{}^{\b} \, (D_{\b} \Psi)\cr
T_{\a \, \b}{}^{\g} \, &=~ \fracm {\,1}{\,2} \, (D_{( \a} \Psi 
)\, \d_{\b)}{}^{\g}  ~+~ \fracm 12 \ell_0 (D_{\d} \Psi )\, (
\g^{[2]})_{( \a}{}^{\d} (\g_{[2]})_{\b )}{}^{\g} \cr 
&{~~~\,}~+~ \ell_1 \, (\g^c)_{\a \, \b} (\g_c)^{\d \, \g} \,
(D_{\d} \Psi) \cr  
&=~  \fracm 1{32} \, \Big[ ~\Big( { 32 \ell_1 -1 - 70 \ell_0 } 
\Big) \, (\g^c)_{\a \b} \, (\g_c)^{\g \d} \cr  
&~~~~\,~~~~~~+~ \fracm 1{2} \Big( {38 \ell_0 + 1} \Big) \, (\g^{
[2]})_{\a \b}  (\g_{[2]})^{\g \d} \cr  
&~~~~\,~~~~~~+~ \fracm 1{120} \Big( {10 \ell_0 - 1} \Big) \, 
(\g^{[5]})_{\a \b}  (\g_{[5]})^{\g \d} ~ \Big] \, (D_{\d} \Psi ) 
~~~.
}\label{eq:Clss-3} \ee
In reaching the second form of the last equation, we used two Fierz 
identities in order to make clear the full content of the equation.   
Demanding that the dimension 1/2 torsions vanish leads to {\em {five}} 
independent conditions
\be \eqalign{
&{~~~~~} 0 ~=~ -1 ~-~ 70 \, \ell_0 ~+~ 32 \,\ell_1 ~~~, ~~~
0  ~=~ 1 ~+~ 38 \, \ell_0  ~~~~, ~~~ \cr
{~~~~~~~~~} 0 \, &=~ 1 ~-~ 10 \, \ell_0  ~~~~,~~~~
0 ~=~ 1 ~+~ 2 \, \ell_0  ~~~\,~~, ~~~~~
0 ~=~ \ell_1 ~+~ 2 \, \ell_0 ~~~,
}\label{eq:Clss-33} \ee
on {\em {two}} constants ($\ell_0$ and $\ell_1$).  The resulting system
is thus overdetermined and inconsistent.  Our argument thus far runs
exactly parallel to the case of 4D, $N$ = 1 supergravity.  

The only consistent solution for completely vanishing dimension 1/2 
torsions in this circumstance is  to also impose the extra condition 
$D_{\a} \Psi ~=~ 0 $.  But unlike the 4D, $N$ = 1 theory discussed
above, this has an additional dire consequence,
\be  \eqalign{
D_{\a} \Psi ~=~ 0 ~~&\to ~~ D_{\a} D_{\b} \Psi ~=~ 0
~~\to ~~ [\, D_{\a} ~,~ D_{\b} \, \} \Psi ~=~ 0 \cr
~~&\to ~~ i \,(\g^c)_{\a \b} \pa_c \Psi ~=~ 0 ~~~~~\,~~~~
\to ~~\pa_c \Psi ~=~ 0 ~~~. ~~~~~~~~
}\label{eq:Clss-2} \ee
This implies that $\Psi$ must be a constant so that (\ref{eq:Clss-1})
reduces to a trivial constant re-scaling of the superframe fields.
Thus eleven dimensional superspace geometry is quite unlike its
bosonic counterpart.  Here we see non-constant supergravity fluctuations 
that are conformally related to the flat superspace {\em {necessarily}}
produce dimension 1/2 and spin-1/2 torsion tensors that are non-vanishing.
At most only {\em {two}} of the five independent structures that occur 
in the dimension 1/2 torsions can be set to zero, if the theory is to 
possess a non-trivial conformal compensator.  This is also seen to be
substantially different from the case of 4D, $N$ = 1.  There it was the 
case that the existence of the chiral compensator still permitted the 
existence of {\em {non-trivial}}  fluctuations that are conformally related 
flat superspace even though all dimension 1/2 torsions vanish.  This is 
{\em {not}} possible for the 11D case since the notion of a chiral superfield 
is non-existent for an 11D theory!  

Within the class of derivatives defined by (\ref{eq:Clss-4}) we
next wish to define one that satisfies the conventional constraints
\be
\nabla_a ~=~ i \fracm 1{32} \, (\g_a)^{\a \b} \, [ \, \nabla_{\a} 
~,~ \nabla_{\b}  \, \} ~~~,~~~ T_{\a ~ [  b c ]} ~-~ \fracm 2{55}
(\g_{b c})_{\a}{}^{\b} \, T_{\b ~ d}{}^d ~=~ 0 
~~~.
\label{eq:Clss2}  \ee
The first of these defines E${}_{a}$ and $\o_{a b c}$ in terms of E${
}_{\a}$ and $\o_{\a b c}$ (c.\ f.\ Eq.\ (\ref{eq:Clss-5X})).  The second 
of these defines the spin-connection $\o_{\a b c}$  in terms of the 
anholonomy (c.\ f.\ Eq.\ (\ref{eq:Clss-6X})).  We can satisfy these 
condition by choosing the constants as
\be 
\ell_0 ~=~ \fracm {1}{10} ~~~,~~~ \ell_1 ~=~ \fracm {1}{4}  
~~~,~~~ \ell_2 ~=~ \fracm 15 ~~~,~~~ \ell_3 ~=~ \fracm 1{160} 
~~~.
\label{eq:Clss1}  \ee
 
Utilizing the fluctuations described by (\ref{eq:Clss-1}) and
(\ref{eq:Clss1}), we calculate the complete commutator algebra
associated with this supergravity covariant derivative to find
$$  \eqalign{
T_{\a \, \b}{}^c \, &=~ i \, (\g^c)_{\a \, \b} ~~~,  \cr 
T_{\a \, \b}{}^{\g} \, &=~ \fracm {3}{\,40}\, (\g^{ d e}
)_{\a \, \b}(\g_{d e})^{\d \, \g} \,(D_{\d} \Psi) ~~~, \cr 
{~~~~~} R_{\a \, \b}{}^{d \, e} \, &=~ - \fracm 1{10} \, 
[~ (\g^{d e})_{\a}{}^{\g} \, (D_{[ \b } D_{\g ]} \Psi) ~+~ 
(\g^{d e})_{\b}{}^{\g} \, (D_{[ \a } D_{\g ]} \Psi)  \cr
&{~~~~~~~~~~~~\,}+~ \fracm 1{8} \, (\g^{b})_{\a \, \b} \, 
(\g_{b}{}^{d e})^{\g \, \d} \, ( D_{\g } D_{\d} \Psi) ~]  
~~~, \cr 
T_{\a \, b}{}^c \, &=~ \fracm {3}{5} \, [~ 2 \, (D_{\a} 
\Psi) \, \d_b {}^c ~-~ \fracm 34 \, (\g^c \, \g_b)_{\a}{}^{
\g} \, (D_{\g} \Psi)  ~] ~~~, \cr
T_{\a \, b}{}^{\g} \, &=~ [~ i\, \fracm 18 \, (D_{[ \a}  
D_{\b ]} \Psi ) \, (\g_b)^{\b \g} ~-~ i\, \fracm 1{\,2,880} 
\, (\g^{[3]})^{\d \e} \, (D_{\d} D_{\e } \Psi )\, (\g_{[3]} 
\, \g_b)_{\a}{}^{\g}   \cr
&{~~~~~~}-~ i\, \fracm 1{\,2,880}  \, (\g^{[3]})^{\d \e} \, 
(D_{\d} D_{\e} \Psi ) \, (\g_b \, \g_{[3]})_{\a}  {}^{\g}  
~-~ \fracm 1{20} \, (\g_b \, \g^c)_{\a}{}^{\g} \, (\pa_c 
\Psi) ~ {~}   \cr
&{~~~~~~}+~ \fracm 7{40} \, (\g^c \, \g_b)_{\a}{}^{\g} \,
(\pa_c \Psi) ~] ~~~,   \cr 
R_{\a \, b}{\,}^{d \, e} \, &=~ [~ - \, i \fracm 1{480} 
\, (\g_b {}^{d e})^{\d \e}\, (D_{[ \a|} D_{|\d |} D_{|\e 
]} \Psi)~-~ \fracm 1{80} \, (\g^c \, \g_b{}^{d e})_{\a}{
}^{\g} \, (\pa_{c} D_{\g} \Psi )  \cr 
&{~~~~~~~}+~ \fracm 15 \, (\g^{d e})_{\a}{}^{\g} \, (\pa_b 
D_{\g} \Psi) ~+~ \fracm 15 \,  \, (\pa^{[d} D_{\a} \Psi) 
\, \d_b {}^{e ]} ~]  ~~~, } $$
\be \eqalign{
T_{a \, b}{}^c \, &=~ \fracm 15 [~ 6 \, (\pa_{[a |} \Psi)
\, \d_{|b]}{}^c ~+~ i \fracm 1{8} \,   (\g_{a b}{}^{c} )^{\a 
\b} \, (D_{\a} D_{\b} \Psi) ~]  ~~~, {~~~~~~~~~~~~~~~~} \cr
T_{a \, b}{}^{\g} \, &=~ - i \, \fracm 14 (\g_{[a | })^{\g 
\, \d} \, ( \pa_{|b]} D_{\d} \Psi ) ~~~,  \cr 
R_{a \, b}{\,}^{d \, e} \, &=~ \fracm 15 [~ (\pa_{[a|} 
\pa^d \Psi) \, \d_{|b]}{}^e ~-~ (\pa_{[a|} \pa^e \Psi) \, 
\d_{|b]} {}^{ d} \cr
&{~~~~~~~}~+~ i \fracm 1{32} \, (\g_{[ a} {}^{d e})^{\a \b} \, 
(\pa_{b ]} \, D_{\a} D_{\b} \Psi ) ~] ~~~. 
}  \label{eq:Clss4}  \ee
These equations emphasize a remarkable fact which is true in all
supergeometries.  The most general supergeometry that is conformally 
related to the flat superspace may be viewed as a geometrical 
description of a scalar superfield.  We have long been aware of 
this fact and have used it previously (\cite{2DN4}) to derive the 
first off-shell description of 2D, $N$ = 4 supergravity.  {\em 
{Within}} 2D theories, this result is even more powerful.  It 
implies that the supergravity constraints for 2D theories are in 
one-to-one correspondence with irreducible scalar superfields and 
thus the supergravity constraints are {\it {totally}} determined 
by the differential equations that define 2D irreducible scalar 
multiplets {\em {and}} {\em {vice}}-{\em {versa}}.

The covariant derivatives in (\ref{eq:Clss-1}) may be ``integrated 
with respect to $\Psi$'' so that in the case of a finite conformal
compensator $\Psi$ we find
\be \eqalign{
\nabla_{\a} & = ~ \Psi^{1/2} \, [ ~ D_{\a} ~+~  \fracm 1{10}\, 
(D_{\g} \, ln\, \Psi) \, (\g^{d e})_{\a}{}^{\g} \, {\cal M}_{d e} ~
]  ~~~, \cr
\nabla_{a} & = ~ \Psi \, [ ~ \pa_{a} ~+~  i \fracm 1{4}\, 
(\g_a )^{\g \d} (D_{\g} \, ln\, \Psi) \, D_{\d} ~+~ \fracm 1{5}\, 
(\pa_b \, ln\, \Psi) \, {\cal M}_{a}{}^{b}  \cr
 & {~~~~~~~~} ~+~ i \fracm 1{160} \, (\g_a {}^{d e})^{\g \d} \, 
(D_{\g} D_{\d} \, ln\, \Psi) \,  {\cal M}_{d e}   \cr
 & {~~~~~~~~} ~+~ i \, \fracm {27}{\,800} \, (\g_a {}^{d e})^{\g \d} 
\, (D_{\g} \, ln\,  \Psi) \, (D_{\d} \, ln\, \Psi) \,  {\cal M}_{d e} ~] 
~~~. } \label{eq:ClssFF} \ee

To summarize the main result in this section, we have shown that under
the assumption that 11D supergravity is separable into a Weyl multiplet
superfield (H${}_{\a}{}^m$) and a conformal compensator superfield ($\Psi$),
Howe's Theorem in the limit of vanish Weyl multiplet forbids the appearance
of the conformal compensator superfield.  This behavior is not congruent
with that (e.\ g.\ the discussion from
(\ref{eq:Clss-1XY}-\ref{eq:Clss-4XX})) found in all previous known cases of
off-shell superfield supergravity.

\section {Traditional Approach to Weyl Symmetry in 11D Superspace}

~~~~Although Howe's 1997 paper \cite{HOW} has described a {\it 
{new}} formalism to realize the presence of Weyl symmetry for 11D
superspace, in fact there is a traditional manner for accomplishing 
this goal.  This traditional approach was initiated by Howe \& Tucker
\cite{HOW1} and then developed by others \cite{barHOW}.  We have even 
been able to extend this  traditional description all the way to 10D, 
$N$ = 1 superspace \cite{Gconf}.  In our recent work \cite{GN01}, we 
began the process of extending this traditional approach to the 11D
theory.   As was shown in \cite{GN01}, the newer approach requires 
that there must exist degrees of freedom in addition to those that 
reside in ${\rm E}_{\a}$.  It is thus a {\it {non}}-{\it {minimal}}
realization of Weyl symmetry in 11D superspace.  The traditional 
approach \cite{HOW1}, does not suffer from this drawback.  So it 
will be the goal of this chapter to extend completely the traditional
formalism for Weyl symmetry in superspace to the 11D case and thereby
establish the  realization of Weyl symmetry in a minimal manner within 
this venue.

In our recent work \cite{GN01}, an analysis of the 11D vielbein degrees of
freedom was performed.  This work implies that a solely conventional
set of constraints\footnote{To our knowledge, this is the first
time this particular set of off-shell constraints for 11D \newline
${~~~~~}$ superspace has been suggested. These are different from
our previous works (e.\ g.\ \cite{GN01}) but \newline ${~~~~~}$ 
but are convenient for many purposes.  The factor of 2/55
in particular leads to `nice' nor- \newline ${~~~~~}$ malizations
in many, many subsequent calculations.} can be chosen as
\be
\eqalign{
 i \,(\g_a)^{\a \b} \, T_{\a \b}{}^b &=~ 32 \d_a \, {}^b 
~~~,~~~ (\g_a)^{\a \b} \, T_{\a \b}{}^{\g} ~=~ 0 ~~~,~~~ 
T_{\a \, [ d e]} ~-~ \fracm 2{55} \, (\g_{d e})_{\a}{}^{\g}  
\, T_{\g b}{}^b ~=~0 ~~~, \cr 
(\g_a)^{\a \b} \, R_{\a \b}{}^{d e} &=~ 0 ~~~,~~~ (\g_{a b}
)^{\a \b} \,T_{\a \b}{}^{b} ~=~ 0 ~~~, ~~~ (\g_{[a b|})^{\a 
\b} \, T_{\a \b}{}_{ | c]} ~=~ 0 ~~~,\cr
(\g_{a b c d e})^{\a \b} \, T_{\a \b}{ }^{e} &=~ 0 ~~~, ~~~
\fracm 1{\,6! \,} \, \e_{\[5\]}{}^{a b 
c d e f} \, (\g_{a b c d e})^{\a \b} \, T_{\a \b
}{}_{f} ~=~ 0  ~~~.
}\label{eq:constrts} \ee
The last four constraints determine the degrees of freedom in
E${}_{\a}{}^{\m}$ that correspond to the elements of the coset
\be
{ {{\rm SL}(32,\IR)} \over {~}{\rm SO}(1,10) \otimes {\rm SO}(1,1)
{~}}
~~~.
\label{eq:coset} \ee
The constraints imply that all 11D supergravity fields are contained 
in two semi-prepotentials; $\Psi$ (conformal compensator) and $H_{\a
}{}^m$ (Weyl supermultiplet) and we will use these constraints in the
following.

The last four constraints of our table may be called ``coset conventional 
constraints\footnote{The 11D coset conventional constraints are closely
related to the constraints discussed in \cite{CGNN}.}'' and in fact
their existence is an almost universal feature of superspace supergravity
theories.  The easiest way to understand this is to consider torus
compactification of the 11D result.  So for example, in four dimensional
$N$-extended supergravity, these coset constraints take the form
\be
{ {{\rm SL}(4N, C )} \over {~}{\rm SO}(1,3) \otimes {\rm SO}(1,1)
 \otimes {\rm U}(N)
{~}}
~~~.
\label{eq:coset1} \ee
for $N \ne 4$ and 
\be
{ {{\rm SL}(16,C )} \over {~}{\rm SO}(1,3) \otimes {\rm SO}(1,1)
 \otimes {\rm SU}(4)
{~}}
~~~.
\label{eq:coset2} \ee
for $N = 4$.
It is perhaps also useful to point out that an example of coset conventional
constraints was discussed \cite{GN85} when we provided the first explicit
solution to constraints for 2D, $N$ = 1 supergravity. In fact, 4D, $N$ = 1
supergravity is the exception rather than the rule when it comes to coset
conventional constraints. While 4D, $N$ = 1  superspace supergravity theory
does not require such constraints, most generic superspace supergravity
theories do require this type of constraint.  

A perennial question we often encounter when discussing superspace
systems of constraints is caused by the fourth entry in our table.
The question is, ``Shouldn't $\o_{a ~ b c}$ be determined by the
condition that appears in (\ref{eq:Clss-4X})?''  The answer to this
is that one can certainly replace the first constraint in the second row 
by $T_{a b c} = 0$.  However, this determines the connection in an
``un-improved manner.''  Certain improvement terms (as seen in the 
4D, $N$ = 1 case) will occur with our choice.

Motivated by the discussion in the last two chapters and by our work 
in \cite{GN01,Gconf}, we can define a {\em  {minimal}} realization of 
a scale transformation law for the 11D superspace covariant derivative 
by
\be \eqalign{ {~~~}
\d_S \nabla_{\a}\, &=~ \fracm 12 L \, \nabla_{\a} ~+~ \fracm {1}{10} 
(\nabla_{\g}  L)  \, (\g^{ b c})_{\a}{}^{\g} {\cal M}_{b c} ~~~,
~~~ \cr
\d_S \nabla_{a}\, &=~ L \, \nabla_{a} ~+~  i\, \fracm {1}{4}\, 
(\g_a)^{\a \, \b} \, (\nabla_{\a} L) \,  \nabla_{\b} 
~+~ \fracm {\,1\,}{\,5\,}  \, (\nabla_c L) \, {\cal M}_{a}{}^c \cr
&{~~~}~+~i \fracm 1{160}\, (\g_{a}^{~b c} )^{\a \b} \, (\nabla_{\a}
\nabla_{\b} L)  \, {\cal M}_{b c}  ~~~. } 
\label{eq:Weyltt} \ee
The importance of these equations is that they permit us to calculate
the scale variations of all superspace torsions and curvatures at the
full non-linear level.  It should also be clearly understood that the
derivatives $\nabla_A$ that appear in (\ref{eq:Weyltt}) are {\em {not}}
restricted to describe fluctuations conformally related to flat
superspace as in (\ref{eq:Clss-1}).  {\em {In particular the torsions in
this chapter are}} {\underline {not}} {\em {given solely by the terms
in}} (\ref{eq:Clss4}).  We find 
$$\eqalign{
\d_S T_{\a \b}{}^c \, &=~ 0 ~~~, \cr
\d_S T_{\a \b}{}^{\g} \, &=~ \fracm 12 L \, T_{\a \b}{}^{\g} ~-~ i 
\fracm 14 \, T_{\a \b}{}^c \, (\g_c)^{\d \g} \,(\nabla_{\d} L) 
~+~\fracm {\,1\,}{2} (\nabla_{( \a} L)\, \d_{\b)}{}^{\g} \cr 
&{~~~}~+~ \fracm {\,1\,}{20}  (\nabla_{\d} L) \, (\g^{[2]})_{(\a}{
}^{\d} \, (\g_{[2]})_{\b)}{}^{\g} ~~~, \cr
\d_S T_{\a b}{}^c \, &=~  \fracm 12 L \, T_{\a b}{}^c ~-~ i \fracm 14 
\, (\g_b )^{\g \d} \, (\nabla_{\g} L) \, T_{\a \d}{}^c ~+~ (\nabla_{
\a} L) \, \d_b {}^c   \cr 
&{~~~}~+~ \fracm 15 \, (\nabla_{\g} L) \, (\g_b {}^c)_{\a}{}^{\g} 
~~~, {~~~~~~~~~~~~~~~~~~~~~~~~~~~~~~}
} $$
\be \eqalign{ {~~~~~}
{~~~~~} \d_S T_{\a b}{}^{\g} \, &=~ L \, T_{\a b}{}^{\g} ~-~ i \fracm 
14 \, (\g_b)^{\d \e} \, (\nabla_{\d} L) \, T_{\a \e}{}^{\g} ~-~
i \fracm 14 \, T_{\a b}{}^c \, (\g_c)^{\d \g} \, (\nabla_{\d} L) \cr 
&{~~~} ~-~ \fracm 12 \, (\nabla_b L) \, \d_{\a}{}^{\g} ~-~ \fracm 
{\,1\,}{10} \, (\nabla_d L) \, (\g_b {}^d)_{\a}{}^{\g}  \cr 
&{~~~} ~+~ i \fracm 14 \, (\g_b)^{\d \g} \, (\nabla_{\a} \nabla_{\d}
L) ~-~ i \fracm 1{320} \, (\g_b {}^{d e})^{\d \e} \, (\nabla_{\d} \nabla_{\e}
L) \, (\g_{d e})_{\a}{}^{\g}  ~~~, \cr
\d_S T_{a b}{}^c \, &=~ L \, T_{a b}{}^c ~+~ i \fracm 14 (\g_{\[a|})^{\d 
\e} \, (\nabla_{\d} L) \, T_{\e |b\]} {}^c
~+~ \fracm 65 \, (\nabla_{\[a|} L )\,
\d_{|b \]}{}^c \cr
&{~~~}~+~ i \fracm 1{40} \, (\g_{a b}{}^c)^{\d \e} (\nabla_{\d} 
\nabla_{\e} L) ~~~,
{~~~~~~~~~~~~~~~~~~~~~~~}{~~~~~~~~~~~~~~~~~~}
\cr
\d_S T_{a b}{}^{\g} \, &=~ \fracm 32 L \, T_{a b}{}^{\g} 
~-~i \fracm 14 \, T_{a b}{}^c \, (\g_c)^{\d \g} \, (\nabla_{\d} L) 
~+~ i \fracm 14 \, (\g_{[a |})^{\a \b} \, (\nabla_{\a} L) \,
T_{\b | b]}{}^{\g} \cr
&{~~~~\,}-~ i \fracm 1{4} \, (\g_{[a|})^{\d \g} (\nabla_{|b]} 
\nabla_{\d} L) ~~~,
}\label{eq:WeylT} \ee
for the various torsion tensor components and 
\be \eqalign{
\d_S R_{\a \b \,}{}^{d e} \, &=~ L \, R_{\a \b \, }{}^{d e} ~+~
\fracm 15 \, T_{\a \b}{}^{[d} \, (\nabla^{e]} L) ~+~
\fracm 15 \, T_{\a \b}{}^{\d} \, (\nabla_{\g} L) \, (\g^{d
e})_{\d}{}^{\g} \cr
&{~~~}~-~ \fracm 15 (\nabla_{(\a |} \nabla_{\g} L) \, (\g^{d
e})_{|\b)}{}^{\g} ~+~ i \fracm {\,1\,}{80} \, T_{\a \b}{}^c (\g_c 
{}^{d e} )^{\d \e} \, (\nabla_{\d} \nabla_{\e} L)
 ~~~, \cr
{~~~} \d_S R_{\a b \,}{}^{d e} \, &=~ \fracm 32 L \, R_{\a b \,}{}^{d 
e} ~-~ i \fracm 14 \, (\g_b)^{\d \e} \, (\nabla_{\d} L) \, R_{\a \e 
\,}{}^{d e} ~+~ \fracm 15 \,T_{\a b}{}^{\g} \, ( \nabla_{\d} L) \, 
(\g^{d e})_{\g}{}^{\d}  \cr 
&{~~~} ~+~ \fracm 15 \, T_{\a b}{}^{[d } \, (\nabla^{e]} L) 
~+~ \fracm 15 \, ( \nabla_{\a} \nabla^{[ d} L) \, \d_b {}^{e]}
~+~ \fracm 15 \, ( \nabla_{b} \nabla_{\g} L) \, (\g^{d e})_{\a} 
{}^{\g}   \cr 
&{~~~} ~+~ i \fracm 1{80} \, T_{\a b}{}^c \, (\g_c {}^{d e})^{\d \e}
\, ( \nabla_{\d} \nabla_{\e} L) ~-~ i \fracm 1{80} \,  (\g_c {}^{b e}
)^{\d \e} \, ( \nabla_{\a} \nabla_{\d} \nabla_{\e} L) 
~~~, \cr
\d_S R_{a b \,}{}^{d e} \, &=~ 2 L \, R_{a b \,}{}^{d e} ~+~ 
i \, \fracm 14 \, (\g_{[a |})^{\a \b} \, (\nabla_{\a} L) \,
R_{\b | b] \,}{}^{ d e } ~+~ \fracm 15 \,T_{a b}{}^{\g} \,  (\g^{d
e})_{\g}{}^{\d}\, ( \nabla_{\d} L) ~~~, \cr
&{~~~~\,}- ~\fracm 15 \, T_{a b}{}^{ [d} \, (\nabla^{e]} L)
~+~ i \, \fracm 1{80} \, T_{a b}{}^{c} \, (\g_c {}^{d e})^{\a \b}
\, (\nabla_{\a } \nabla_{\b } L) \cr
&{~~~~\,}+ ~\fracm 15 \,  \delta_{ [a |}{}^{[d }
 \, ( \nabla_{b]} \nabla^{e]} L) ~+~ i \, \fracm 1{80} \, (\g^{d e}{}_{ [
a})^{\a \b}
\, (\nabla_{b] } \nabla_{\a } \nabla_{\b } L) ~~~,
}\label{eq:WeylR} \ee
for the various curvature tensor components.

Although these conformal transformation laws may seem quite
complicated when compared to those of (\cite{HOW}), an investigation of
previous lower dimensional off-shell supergravity theories will reveal
many similarities.

An especially important superfield occurs at dimension 1/2.  We denote 
this superfield by ${\cal J}_{\a}$ where
\be \eqalign{ 
{\cal J}_{\a} ~\equiv~ \fracm {4}{33} \, T_{\a b}{}^b  ~~ 
\to ~~ \d_S {\cal J}_{\a} ~=~ \fracm 12 L \, {\cal J}_{\a} ~+~ 
(\nabla_{\a} L) ~~~. \cr
} \label{eq:Weyl01} \ee
This transformation law implies that ${\cal J}_{\a}$ is {\em {not}}
a superconformal invariant superfield.

It has been proposed \cite{CGNN} that a rank six and dimension zero
field strength denoted by $X_{[5]}{}^a$ is critical to obtain
a  superspace description of the M-theory effective action.  In our 
previous work \cite{GN01}, we pointed out that there is also an 
alternate possibility of allowing another similar tensor, $X_{[2]}{}^a$ . 
These both can be found in 
\be
T_{\a \b}{}^c ~=~ i \,(\g^c)_{\a \b} ~+~  \,\fracm 12 \,  (\g^{
\[2\]})_{\a \b} \, X_{\[2\]}{}^c ~ +~ i\, \fracm 1{120} \, 
(\g^{\[5\]})_{\a \b} \, X_{\[5\]}{}^c  
\label{eq:XT} ~~~, \ee
or alternately we see
\be 
\eqalign{
 X_{\[a b \]}{}^k &\equiv~ \fracm 1{32} \, (\g_{a b})^{\a \b} \, 
T_{\a \b}{}^k ~~~,
\cr X_{\[ a b c d e\]}{}^k  &\equiv~ i \, \fracm 1{32} \, (\g_{a b 
c d e})^{\a \b} \, T_{\a \b}{}^k ~~~.
} \label{eq:XT1}  \ee

It is interesting to note what result is obtained from the quantity
defined by $T_{\a b}{}^b - (3/4) T_{\a \b}{}^{\b}$ under the action of the
scale transformation law in (\ref{eq:Weyltt}).  In fact this object is
a dimension 1/2 spin-1/2 scale covariant.
Starting from
the third equation in (\ref{eq:WeylT}) we see
\be  \eqalign{ 
\d_S T_{\a \b}{}^{\b} \, \, &=~ \fracm 12 L \, T_{\a \b}{}^{\b} ~-~ 
i \fracm 14 \, T_{\a \b}{}^c \, (\g_c)^{\d \b} \,(\nabla_{\d} L) 
~+~\fracm {\,1\,}{2} (\nabla_{( \a} L)\, \d_{\b)}{}^{\b} \cr 
&{~~~}~+~ \fracm {\,1\,}{20}  (\nabla_{\d} L) \, (\g^{[2]})_{(\a}{
}^{\d} \, (\g_{[2]})_{\b)}{}^{\b} \cr 
&=~ \fracm 12 L \, T_{\a \b}{}^{\b} ~-~ i \fracm 14 \, [ \, i (\g^c
)_{\a \b} ~+~  \,\fracm 12 \,  (\g^{\[2\]})_{\a \b} \, X_{\[2\]}{}^c  
\cr 
&{~~~}~ +~ i\, \fracm 1{120} \, (\g^{\[5\]})_{\a \b} \, X_{\[5\]}
{}^c  ~] \, (\g_c)^{\d \b} \,(\nabla_{\d} L)  \cr 
&{~~~}~+~\fracm {\,1\,}{2} (\nabla_{( \a} L)\, \d_{\b)}{}^{\b} ~+~ 
\fracm {\,1\,}{20}  (\nabla_{\d} L) \, (\g^{[2]})_{(\a}{ }^{\d} \,
(\g_{[2]})_{\b)}{}^{\b} \cr
&=~ \fracm 12 L \, T_{\a \b}{}^{\b} ~-~  \fracm {11}4 \, 
\,(\nabla_{\a} L)  ~+~\fracm {33}{2} (\nabla_{\a} L) ~-~ 
\fracm {55}{20} (\nabla_{\a} L)  ~~~,  \cr
~~~\to~~ \d_S ( \fracm 1{11} T_{\a \b}{}^{\b} ) \, &=~ \fracm 12 L \, 
(\fracm 1{11} \, T_{\a \b}{}^{\b}) ~+~  (\nabla_{\a} L) ~~~.
} \label{eq:Weyl010} \ee
The reason the terms involving the $X$-tensors have ``disappeared''
from the final result can be seen from the following considerations.
\be \eqalign{
\fracm 12 \,  (\g^{\[2\]})_{\a \b} \, (\g_c)^{\d \b} \, X_{\[2\]
}{}^c &=~ - \, \fracm 12 \,  (\g_{\[2\]} \, \g_c)_{\a} {}^{\d} \,
X^{\[2\] }{}^c \cr 
&=~ - \, \fracm 12 \, [ \, (\g_{\[ a b c\]})_{\a} {}^{\d} ~-~ 2 
\eta_{c a} (\g_{b})_{\a} {}^{\d} \,] \, \, X^{\[a b\]}{}^c  ~
=~ 0 ~~~, \cr
{~~~~} i  \fracm 1{120}  \,  (\g^{\[5\]})_{\a \b} \, (\g_c)^{\d \b} 
\, X_{\[5\] }{}^c &=~ - i  \fracm 1{120}  \,  (\g_{\[5\]} \, \g_c
)_{\a}{}^{\d} \, X^{\[5\] }{}^c \cr 
&=~  \fracm 1{120} \, [ \,  \fracm 1{120} \, \e_{a b c d e f}{}^{\[
5^{\prime}\]} (\g_{\[ 5^{\prime} \]})_{\a} {}^{\d}  \cr 
&~~~~~~~~~~~~-~ i 15 \eta_{c a} (\g_{b d e f})_{\a} {}^{\d} \,] \, \,
X^{\[a b d e f\]}{}^c  ~  =~ 0 ~~~.
}\label{eq:WeyX10sR} \ee
The conventional constraints imposed upon the $X$-tensors eliminate
their appearance from the final result in (\ref{eq:Weyl010}). So that
we see an alternate definition of the ${\cal J}$-tensor is given by
${\cal J}_{\a} = \fracm 1{11}T_{\a \b}{}^{\b}$.  The appearance of
this non-scalar covariant field Poincar\' e supergravity strength 
should not come as a surprise.  At least one non-scale invariant field
strength superfield can be verified in every known off-shell 
superspace formulation that has ever been given.

The reason that this superfield is important is that it allows the
definition of a Poincar\' e superspace spinorial differentiation operation
that  acts consistently to remain solely within the space of superconformal
tensors.  To see this, we note that a superconformal tensor denoted 
by ${\cal T}^{(w)}{}_{a_1 \dots a_p}{}^{b_1 \dots b_q}$ and of weight
$w$ can be defined to transform according to
\be
{~~~}
\d_S {\cal T}^{(w)}{}_{a_1 \dots a_p}{}^{b_1 \dots b_q}  ~=~ 
w\, L\, {\cal T}^{(w)}{}_{a_1 \dots a_p}{}^{b_1 \dots b_q} ~~~,
 \label{eq:conTENS1} \ee
under the action of the superspace scale transformation.  From this
it follows that we find
\be \eqalign{ {~~~~~~}
\d_S (\, \nabla_{\a} {\cal T}^{(w)}{}_{a_1 \dots a_p}{}^{b_1 \dots b_q} 
\,) &=~ ( w + \fracm 12) \, L \, (\nabla_{\a} {\cal T}^{(w)}{}_{a_1 \dots
a_p}{}^{b_1 \dots b_q}) \cr
&{~~~~} ~+~ w\, (\nabla_{\a} L) \,  {\cal T}^{(w)}{}_{a_1 \dots 
a_p}{}^{b_1 \dots b_q}  \cr
&{~~~~} ~+~  \fracm {1}{10} (\nabla_{\g}  L)  \, (\g^{ b c})_{\a}{}^{\g} 
(\, {\cal M}_{b c} \, {\cal T}^{(w)}{}_{a_1 \dots a_p}{}^{b_1 \dots b_q}
\,) ~~~, \cr
\d_S (\, {\cal J}_{\a} ~{\cal T}^{(w)}{}_{a_1 \dots a_p}{}^{b_1 \dots b_q} 
\,) &=~ ( w + \fracm 12) \, L \, (~{\cal J}_{\a} ~ {\cal T}^{(w)}{}_{a_1 
\dots a_p}{}^{b_1 \dots b_q}) \cr
&{~~~~} ~+~  (\nabla_{\a} L) \,  {\cal T}^{(w)}{}_{a_1 \dots 
a_p}{}^{b_1 \dots b_q}  \cr 
\d_S (\, {\cal J}_{\g}  \, (\g^{ b c})_{\a}{}^{\g}  {\cal M}_{b c} ~{\cal
T}^{(w)}{}_{a_1 \dots a_p}{}^{b_1 \dots b_q} 
\,) &=~ ( w + \fracm 12) \, L \,{\cal J}_{\g}  \, (\g^{ b c})_{\a}{}^{\g} 
~ (~ {\cal M}_{b c} \,{\cal T}^{(w)}{}_{a_1 
\dots a_p}{}^{b_1 \dots b_q}) \cr
&{~~~~} ~+~  (\nabla_{\g} L)  \, (\g^{ b c})_{\a}{}^{\g} \,  ({\cal M}_{
b c} \, {\cal T}^{(w)}{}_{a_1 \dots  a_p}{}^{b_1 \dots b_q} ) ~~~. 
} \label{eq:conTENS2} \ee
These three equation taken together inform us that the quantity defined by
\be
( {\Hat {\nabla}}{}_{\a} {\cal T}^{(w)}{}_{a_1 \dots a_p}{}^{b_1 \dots b_q} )
~\equiv~ [~ ({\nabla}{}_{\a} ~-~ w \, {\cal J}{}_{\a}
~-~ \fracm 1{10} {\cal J}_{\g}  \, (\g^{ b c})_{\a}{}^{\g}  {\cal M}_{b c} )
\, {\cal T}^{(w)}{}_{a_1 \dots a_p}{}^{b_1 \dots b_q} ] ~~~,
\label{eq:conTENS3} \ee
possesses a covariant scale transformation law
\be \eqalign{ {~~~~~~}
\d_S (\, {\Hat {\nabla}}{}_{\a} {\cal T}^{(w)}{}_{a_1 \dots a_p}{}^{b_1 
\dots b_q} \,) &=~ ( w + \fracm 12) \, L \, ({\Hat {\nabla}}{}_{\a}   
{\cal T}^{(w)}{}_{a_1 \dots a_p}{}^{b_1 \dots b_q}) ~~~,
}\label{eq:conTENS4} \ee
with scale weight $( w + \fracm 12)$.  

Examples of 11D scale-covariant supergravity tensors are provided by 
the ``$X$-tensors'' \cite{CGNN}\footnote{In their work, Cederwall, Gran,
Nielsen and Nilsson only retained the $X_{[5]}{}^a$ tensor.}  According 
to the first result in (\ref{eq:WeylT}) both $X$-tensors are scale- 
covariant having $w = 0$.  Thus, we see
\be  \eqalign{
{~~~~~~}
\d_S \, X_{\[2\]}{}^c \, &=~ 0 ~~~, ~~~ \d_S \, 
 X_{\[5\]}{}^c  ~=~ 0 ~~~, ~~~\to  \cr
{\Hat {\nabla}}{}_{\a} X_{\[2\]}{}^c \, & = ~ ({\nabla}{}_{\a} ~-~ 
\fracm 1{10} {\cal J}_{\g}  \, (\g^{ b c})_{\a}{}^{\g} {\cal M}_{
b c} ) \, X_{\[2\]}{}^c ~~~,\cr
~~~ {\Hat {\nabla}}{}_{\a} X_{\[5\]}{}^c \, &=~ ({\nabla}{}_{\a} 
~-~ \fracm 1{10} {\cal J}_{\g} \, (\g^{ b c})_{\a}{}^{\g} {\cal 
M}_{b c} ) \, X_{\[5\]}{}^c ~~~.
}\label{eq:XT3}  \ee
and consequently
\be \eqalign{ {~~~~~}
\d_S (\, {\Hat {\nabla}}{}_{\a} X_{\[2\]}{}^c \,) ~=~ \fracm 12 
\, L \, ({\Hat {\nabla}}{}_{\a} X_{\[2\]}{}^c ) ~~~, ~~~ \d_S (\, 
{\Hat {\nabla}}{}_{\a} X_{\[5\]}{}^c \,) ~=~ \fracm  12 \, L \,
({\Hat {\nabla}}{}_{\a} X_{\[5\]}{}^c ) ~~~, }
\label{eq:XT4}  \ee
and we find the ``hatted derivative'' of the $w = 0$ scale-covariant 
$X$-tensors are $w = 1/2 $ scale-covariant tensors.   All of these 
results are the expected generalization of the ones found for the 
10D, $N$ = 1 superspace \cite{Gconf}.

A second especially important superfield occurs at dimension one 
and is the on-shell 11D supergravity field strength.  We denote this
superfield by ${\cal  W}_{a b c d}$ and note that it contains the 
usual supercovariantized Weyl tensor\footnote{We use the non-calligraphic
symbol $W_{a b c d}$ for the usual Weyl tensor superfield.} at second 
order in its $\q$ expansion.   Given the superspace scale transformation 
law of the various torsion components (\ref{eq:WeylT}), it follows 
that a superscale-covariant quantity of dimension one and with $w = 
1$ is given by
\be \eqalign{ {~~}
{\cal W}_{a b c d} &\equiv~ \fracm 1{32}\, \Big[\, i \,( \g^e \g_{a 
b c d})_{\g}{}^{\a} \, T_{\a \, e}{}^{\g} ~-~ \fracm {1}3 \, (\g_{
a b c d})^{\a \b} ( \,  \nabla_{\a} T_{\b c}{\,}^c \,+\, 
\fracm {14}{1,815} \,  T_{\a c} {\,}^c \, \, 
T_{\b d}{\,}^d  \,) ~ \Big] \cr
&~~~\to ~ \d_S \, {\cal W}_{a b c d}  ~=~ L \, {\cal W}_{a b c d} ~~~.
} \label{eq:onshellF} \ee
This 4-form superfield quite properly may be called the 11D Weyl 
multiplet superfield.  However, as we noted above, the component-level
supercovariantized Weyl tensor is {\em {not}} the leading component
of this superfield.  Since this superfield contains the Weyl multiplet, 
it does not vanish when the 11D supergravity multiplet obeys the equations
of motion associated with 11D super Poincar\' e action.  It also thus
corresponds to the on-shell field strength of 11D supergravity. Our
discussion above  can be applied to solve a small puzzle in the work of
\cite{CGNN}.   Namely this formula necessarily defines the on-shell field
strength,  containing the physical degrees of freedom of the 11D theory, 
as it picks out a particular linear combination of the two four-forms 
described in the work of Ref.\ \cite{CGNN}\footnote{The definition of ${\cal W
}_{a b c d}$ can change slightly depending on the choice of conventional 
\newline ${~~~~~}$ constraints.  Our definition is the one that follows 
from choosing the constraints as \newline ${~~~~~\,}$ in
(\ref{eq:constrts}).}. 

The equation in (\ref{eq:conTENS3}) has an important component implication.
If we denote the local supersymmetry variation in the putative 11D
superconformal theory by ${\Hat \d}{}_Q$, the corresponding one in
the 11D Poincar\' e theory by $\d_Q$, the local scale variation by $\d_S$
and the local Lorentz variation by $\d_{LL}$, then Eq.\ (\ref{eq:conTENS3})
implies
\be
{\Hat \d}{}_Q(\e^{\a}) ~=~ \d_Q(\e^{\a}) ~-~ \d_S(\e^{\a} \l_{\a}) 
~-~ \fracm 1{10} \d_{LL}(\e^{\a} (\g^{ b c})_{\a}{}^{\b} \l_{\b})  ~~~,
\label{eq:conTENS44} \ee
where $\e^{\a}(x)$ is the local supersymmetry parameter and $\l_{\a}
(x)$ is the $\q \to 0$ limit of ${\cal J}_{\a}$.

The main results of this section are summarized in the following table.
\vspace{0.1cm}
\begin{center}
\footnotesize
\begin{tabular}{|c|c|c|c|}\hline
${\rm {~~~Superfield~~~}} $ & ${\rm {~~Conformal~~}}$ & ${\rm
{~~Weight~~}}
$  & ${\rm {~~Engineering ~dim.~~}} $ \\
\hline
$ ~{\cal W}_{a b c d}~$ & $ ~{\rm {Yes}}~$ & $ ~ 1 ~$  & $ ~ 1 ~$ \\
\hline
$ ~{\cal J}_{\a}~$ & $ ~{\rm {No}}~$ & $ ~1/2~$  & $ ~ 1/2 ~$ \\ \hline
$ ~X_{\[a b d e f\]}{}^c~$ & $ ~{\rm {Yes}}~$ & $ ~ 0~$  & $ ~ 0 ~$ \\
\hline
$ ~X_{\[a b \]}{}^c~$ & $ ~{\rm {Yes}}~$ & $ ~0~$ & $ ~ 0 ~$ 
\\ \hline
\end{tabular}
\end{center}
\begin{center}
{Table 1:  11D Supergravity Field Strength Superfields}
\end{center}

\section{Weyl 11D Superspace Geometry from Super\\ Poincar\' e Geometry}

~~~~The spinorial ``hatted'' supercovariant derivative introduced 
in the last chapter allows the construction of supergeometrical objects
that possess superconformal symmetry.  The spinorial Weyl ``hatted'' 
supercovariant derivative given in (\ref{eq:conTENS3})
\be
{\Hat {\nabla}}{}_{\a} ~\equiv~ {\nabla}{}_{\a} ~-~ w \, {\cal 
J}{}_{\a} ~-~ \fracm 1{10}  {\cal J}_{\g}  \, (\g^{ b c})_{\a
}{}^{\g} {\cal M}_{b c}  ~~~,
\label{eq:conTENS3B} \ee
may be used to define a bosonic Weyl ``hatted'' supercovariant derivative,
\be \eqalign{ {~~}
{\Hat {\nabla}}{}_a &\equiv~ i \fracm 1{32} \, (\g_a)^{\a \b} 
\, [ \, {\Hat {\nabla}}{}_{\a} ~,~ {\Hat {\nabla}}{}_{\b} \, 
\} ~~~, \cr
&=~ \nabla_a ~-~ i \, \fracm 14 \, (\g_a)^{\a \b} \, {\cal J
}_{\a} \nabla_{\b} ~-~ i w \, \fracm 1{16} \,(\g_a)^{\a \b} 
\, [\, \nabla_{\a}{\cal J}_{\b} \, ] \,{\bf I}  \cr
&~~~~
~-~ i \, \fracm 1{160} \,  (\g_a \g_{d e}){}^{\a \b} \, [\,  
({\nabla}_{\a} \,-\, \fracm {13}5 \, {\cal J}_{\a}) \,{\cal 
J}_{\b} \,] \, {\cal M}^{d e}   ~~~.
}\label{eq:Clss2B}  \ee
From the results in (\ref{eq:Weyltt}), (\ref{eq:Weyl01}) and 
(\ref{eq:conTENS4}) we find the following variations,
\be
\eqalign{ {~}
\d_S [ \, \nabla_a {\cal T}^{(w)}\,] &=~ (w + 1 )\, L \nabla_a {\cal
T}^{(w)} \,+\, i \fracm 14 \, (\g_a)^{\a \b} \, (\nabla_{\a} L)
(\nabla_{\b} {\cal T}^{(w)}) \,+\, w \, (\nabla_a L) \,{\cal T}^{(w)} ~\cr
&{~~~~}+~ \fracm 15  \,
(\nabla_d L) \,({\cal M}_a {}^d {\cal T}^{(w)} )  ~+~ i \fracm 1{160} \,
(\g_a {}^{d e})^{\a \b} \,  (\nabla_{\a} \nabla_{\b} L) \, ({\cal M}_{d e}
{\cal T}^{(w)} ) ~~~,
\cr }\label{eq:confID1} 
\ee
\be
\eqalign{ {~~}
\d_S [ \, i \, (\g_a)^{\a \b} \, {\cal J }_{\a} (\nabla_{\b}
{\cal T}^{(w)}) \,] &=~ i (w + 1 ) L  \, (\g_a)^{\a \b} \, 
{\cal J }_{\a} (\nabla_{\b} {\cal T}^{(w)})  \cr
&{~~~~\,}+~ i  \, (\g_a)^{\a \b} \, (\nabla_{\a} L)
(\nabla_{\b} {\cal T}^{(w)})   \cr
&{~~~~\,}+~ i w  \, (\g_a)^{\a \b} \, {\cal J}_{\a} (\nabla_{\b} L)
{\cal T}^{(w)}   \cr
&{~~~~}+~ i \fracm 1{10} \, (\g_a {\g}^{d e})^{\a \b} \, 
{\cal J}_{\a} (\nabla_{\b} L) \, ({\cal M}_{d e} {\cal T}^{(w)} ) 
~~~,
\cr }\label{eq:confID2}  \ee
\be
\eqalign{ {~~}
\d_S [ \, i \, (\g_a)^{\a \b} \, ( \nabla_{\a} {\cal J }_{\b})
\,{\cal T}^{(w)}  \,] &=~ i (w + 1 ) L  \, (\g_a)^{\a \b} \, ( 
\nabla_{\a} {\cal J }_{\b}) \,{\cal T}^{(w)}  
\,+\, 16 (\nabla_a L) \,{\cal T}^{(w)}  \cr
&{~~~~\,}-~ i 4  \, (\g_a)^{\a \b} \, {\cal J}_{\a} (\nabla_{\b} L)
{\cal T}^{(w)} ~~~, \cr
}\label{eq:confID3}  \ee
\be
\eqalign{ {~~}
\d_S [ \, i \, (\g_d)^{\a \b} \, ( \nabla_{\a} {\cal J }_{\b})
({\cal M}_a {}^d {\cal T}^{(w)}) \,] &=~ i (w + 1 ) L  \, (\g_d)^{
\a \b} \, ( \nabla_{\a} {\cal J }_{\b}) ({\cal M}_a {}^d {\cal 
T}^{(w)})   \cr
&{~~~~\,}-\, i 4 (\g_a)^{\a \b} \, {\cal J}_{\a} (\nabla_{\b} L)
\,({\cal M}_a{}^d {\cal T}^{(w)}) 
\cr
&{~~~~\,}+\, 16 (\nabla_d L) \,({\cal M}_a{}^d {\cal T}^{(w)}) 
~~~, \cr }\label{eq:confID4}  \ee
\be
\eqalign{ {~~}
\d_S [ \, i \, (\g_a {}^{d e})^{\a \b} \, ( \nabla_{\a} {\cal J 
}_{\b}) ({\cal M}_{d e} {\cal T}^{(w)} ) \,] &=~ i \, (w + 1 ) L 
\, (\g_a {}^{d e})^{\a \b} \, ( \nabla_{\a} {\cal J 
}_{\b}) ({\cal M}_{d e} {\cal T}^{(w)} )   \cr
&{~~~~\,}+\, i \fracm 65 (\g_a{}^{d e})^{\a \b} \, {\cal J}_{\a}
(\nabla_{\b} L)\,({\cal M}_{d e} {\cal T}^{(w)})   \cr
&{~~~~\,}+\, i \, (\g_a {}^{d e})^{\a \b} \,  (\nabla_{\a} \nabla_{\b} L)
\, ({\cal M}_{d e} {\cal T}^{(w)} )~~~, \cr
}\label{eq:confID5}  \ee
\be
\eqalign{ {~~}
\d_S [ \, i \, (\g_a {}^{d e})^{\a \b} \, ( {\cal J}_{\a} {\cal 
J }_{\b}) ({\cal M}_{d e} {\cal T}^{(w)}) \,] &=~ i (w + 1 ) L 
\, (\g_a {}^{d e})^{\a \b} \, ( {\cal J}_{\a} {\cal 
J }_{\b}) ({\cal M}_{d e} {\cal T}^{(w)})    \cr
&{~~~~\,}+\, i 2 (\g_a{}^{d e})^{\a \b} \, {\cal J}_{\a}
(\nabla_{\b} L)\,({\cal M}_{d e} {\cal T}^{(w)}) ~~~, \cr
}\label{eq:confID6}  \ee
In writing these we have used the short-hand notation $ {\cal
T}^{(w)} \equiv {\cal T}^{(w)}{}_{a_1 \dots a_p}{}^{b_1 \dots b_q}$.
The definition of ${\Hat {\nabla}}{}_a $ insures that the following
relation is satisfied,
\be
\eqalign{ {~~~}
\d_S {\cal T}^{(w)}{}_{a_1 \dots a_p}{}^{b_1 \dots b_q}  &=~ w\, 
L\, {\cal T}^{(w)}{}_{a_1 \dots a_p}{}^{b_1 \dots b_q} ~\to \cr
&\d_S ( {\Hat {\nabla}}{}_a{\cal T}^{(w)}{}_{a_1 \dots a_p}{}^{b_1 
\dots b_q} \,) ~=~ ( w + 1 )\, L\, {\Hat {\nabla}}{}_a{\cal T}^{
(w)}{}_{a_1 \dots a_p}{}^{b_1 \dots b_q} ~~~. } 
\label{eq:conTENS1B} \ee

A calculation of the graded commutator algebra of the ``hatted''
derivatives yield ``hatted'' torsion and curvature superfields.
As well a dilatation field strength superfield $ {\cal F}_{{\un A} 
\,{\un B}}$ can be defined from the graded commutator,
\be \eqalign{
[\, {\Hat {\nabla}}{}_{\un A} ~,~ {\Hat {\nabla}}{}_{\un B}\, \} 
~\equiv~ & {\Hat T}{}_{\un A \un B}{}^{\g} {\Hat {\nabla}}{
}_{\g} ~+~ {\Hat T}{}_{\un A \un B}{}^{c} {\Hat {\nabla}}{
}_{c} ~+~ \fracm 12 {\Hat R}{}_{\un A \un B}{}_{d }{}^e {\cal 
M}_e {}^d  \cr 
&-~ w \, {\cal F}_{{\un A} \,{\un B}} ~ {\bf I}  ~~~. } 
\label{eq:Clss2C}  \ee
Given the results of (\ref{eq:conTENS3B}) and (\ref{eq:Clss2B}) 
we also find
\be \eqalign{
[\, {\Hat {\nabla}}{}_{\un A} ~,~ {\Hat {\nabla}}{}_{\un B}\, \} 
~=~ &[\,{\nabla}{}_{\un A} ~,~ {\nabla}{}_{\un B}\, \} ~+~ {\cal
K}{}_{\un A \un B}{}^{\un C} {\nabla}{}_{\un C} ~+~ \fracm 12 
{\cal K}{}_{\un A \un B}{}_{d}{}^{e} {\cal M}_{e}{}^{d}  \cr 
&-~ w \, {\cal F}_{{\un A} \,{\un B}} ~ {\bf I}  ~~~.
} \label{eq:Clss2Cc}  \ee
The quantities $ {\cal K}{}_{\un A \un B}{}^{\un C}$, $ {\cal 
K}{}_{\un A \un B}{}^{d e}$ and ${\cal F}_{{\un A} \,{\un B}}$ 
are expressed\footnote{Although the calculation of the ${\cal
K}$-quantities, ${\Hat {\cal K}}$-quantities and as well
${\cal F}_{{\un A} \,{\un B}}$ is straightforward,
\newline ${~~~~~}$ we will forego giving explicit expressions for
all of these.} in terms of the Poincar\' e  superspace covariant 
derivative $ {\nabla}{}_{\un A}$ and $ {\cal J}_{\a}$.   The two equations immediately above inform us that the
super-Weyl  covariants ${\Hat T}{}_{\un A \un B}{}^{\un C}$ and $ {\Hat R
}{}_{\un A \un B}{}_{d }{}^e $ are related to the corresponding 
Poincar\' e objects $ {T}{}_{\un A \un B}{}^{\un C}$ and $ {R}
{}_{\un A \un B}{}_{d  }{}^e $ via,
\be \eqalign{
{\Hat T}{}_{\un A \un B}{}^{\un C} &=~ {T}{}_{\un A \un B}{
}^{\un C}  ~+~  {\Hat {\cal K}}{}_{\un A \un B}{}^{\un C} ~~~,  ~~~
{\Hat R}{}_{\un A \un B}{}_{d }{}^e ~=~ {R}{}_{\un A \un B}{}_{d
}{}^e ~+~ {\Hat {\cal K}}{}_{\un A \un B}{}_{d}{}^{e} ~~~.
}\ee

\section{Scale vs. Non-Scale Invariant Equations of \\Motion}

~~~~The analysis of the conformal properties of the full non-linear 
theory given in a previous chapters once again provides an argument against
the  result known as ``Howe's Theorem'' that asserts that there cannot
appear a dimension 1/2 and spin-1/2 auxiliary field strength superfield. 
Thus  Howe's Theorem is equivalent to the imposition of the condition 
that ${\cal J}{}_{\a} = 0$ be imposed as a kinematic constraint.  Let 
us now impose this  condition in addition to those that appear in 
(\ref{eq:constrts}).   The scale transformation properties of the
superspace are still described by (\ref{eq:Weyltt}).  In turn this
observation together with a consistency condition applied to
(\ref{eq:Weyl01}) leads to the condition $\nabla_{\a}L = 0$. This is 
the full nonlinear extension of (\ref{eq:Clss-2}) and leads to the 
exact same problem as we found in our pre-potential analysis.

Let us use these insights in a different way.  In this very short
section, we will simply compare properties of the non-supersymmetric
Poincar\' e gravity equations of motions with two possible supersymmetric
extensions.

For the ordinary 11D $x$-space covariant derivative, a scale 
transformation law is given by
\be 
\d_S \nabla_a ~=~ \ell \, \nabla_a ~-~ (\nabla_k \ell) {\cal M}_a {}^k
~~~,
\label{eq:compar1} \ee
where $\ell(x)$ is a local scale parameter.  This implies that the 
11D $x$-space Riemann curvature tensor transforms as
\be 
\d_S r_{a  b} {~}^{d e} ~=~ \ell \, r_{a  b} {~}^{d e} ~-~ 
(\nabla_{[a}  \nabla^{[d}  \ell)\, \d_{b]}{}^{e]}
~~~.
\label{eq:compar2} \ee
The constraint is, of course, that the torsion tensor vanishes $t_{a b c}
= 0$.   In the absence of matter, the usual Einstein-Hilbert action leads 
to the expected equation of motion 
\be
{\cal E}_{a b} ~\equiv~ r_{a  c ~ b} {}^{c} ~-~ \fracm 12 \eta_{a b} \,
r_{c  d} {~}^{c d} ~=~ 0
~~~,
\label{eq:compar3} \ee
i.e. the Einstein tensor vanishes.  It is a simple matter to show that
this equation of motion is {\em {not}} a scale-invariant condition
since,
\be
\d_S {\cal E}_{a b} ~=~ 2 \ell \, {\cal E}_{a b} ~-~ 9 \,
(\nabla_{a}  \nabla_{b}  \ell) ~+~ 9 \, \eta_{a b} \,
(\nabla^c  \nabla_c  \ell) 
~~~.
\label{eq:compar4} \ee

Now in the scheme that was proposed in the works in \cite{GN01} and
\cite{G6}, the equations of motions are
\be
{\cal J}_{\a} ~=~ 0 ~~,~~ X_{\[a b \]}{}^c ~=~ 0 ~~,~~
X_{\[a b d e f\]}{}^c ~=~ 0 ~~~,
\label{eq:compar5} \ee
Due to the observation in (\ref{eq:Weyl01}), we see that the first of 
these superfield equations is also {\em {not}} a super scale-invariant 
condition.

This is to be contrasted with the schemes (\cite{HOW}  and \cite{CGNN})
where the equations of motions are
\be
X_{\[a b \]}{}^c ~=~ 0 ~~~,~~~ X_{\[a b d e f\]}{}^c ~=~ 0 ~~~.
\label{eq:compar6} \ee
but the condition $ {\cal J}_{\a} = 0 $ is considered to be a constraint.
Thus, in these schemes the equations of motion {\em {are}} super 
scale covariant.  From this point of view, the role of the
${\cal J}$-tensor is to extend the non-scale covariance of the
Poincar\' e equations of motion in $x$-space into the non-scale covariance
of Poincar\' e equations of motion in superspace.   So to accept the
validity of Howe's Theorem\footnote{The authors of \cite{CGNN}
have noted the need to check explicit M-theory currents for
consistency \newline ${~~~~\,}$ with Howe's theorem.} is equivalent 
to imposing super scale covariance on the equations of motion.

In M-theory, the low-energy effective action is one whose lowest
order terms describe 11D Poincar\' e supergravity.  Some of the
structure of the next order corrections in an expansion $\ell_{11}$
have been discussed in the literature.  None to date lead to
scale-invariant terms appearing in the equations of motion.  Thus, 
it is our position that this alone precludes any superspace obeying 
Howe's Theorem from being able to describe the M-theory effective 
action.  The non-scale covariance of the M-theory effective action
equations of motion demand the appearance of the ${\cal J}$-tensor.

\section{Summary Discussion }

~~~~Under the impetus of the effective M-theory action's superspace
formulation, problems in 11D supergravity are now being faced.  This
is welcomed activity.  However, we are still without truly fundamental
insight into off-shell 11D, $N$ = 1 supergravity, despite the work 
underway.  It is useful to use 4D, $N$ = 1 supergravity to make this 
point most sharply.  We can do this by giving further consideration 
to the situation of the scale compensator.  There is another layer 
to the structure of the theory.  The truly irreducible minimal 4D, 
$N$ = 1 off-shell supergravity compensator satisfies the equation $
{\cal J}_{\a}  = 0$.   The solution to this {\it {in}} {\it {four}} 
{\it {dimensions}} implies that $\Psi$ is a linear combination of a
chiral superfield and its conjugate.  It is the analog of this
irreducibility condition that we still lack in the 11D theory because
chiral superfields cannot exist  in 11D superspace.  

{\it {Since chirality does not exist in 11D, there are no non-trivial 
conformal fluctuations of the flat metric to the equation}} ${\cal 
J}_{\a} = 0$.  

The supergravity vielbein in (\ref{eq:Clss-1}) with its $2^{32}$ 
d.\ f.\  (d.\ f.\ $\equiv$ degrees of freedom) is an off-shell but 
highly reducible supergravity representation.   It requires the 
imposition of differential equations upon it in order to reach a 
{\it {minimal}} {\it {irreducible}} off-shell representation.  Based 
on some structures observed in an unusual class of algebras \cite{GR}, 
we conjecture that  such irreducible representation exist.  For example 
we have found some evidence that there  may well exist a 32,768 bosonic 
d.\ f.\ and 32,768 fermionic  d.\ f.\ irreducible representation of 
D = 11 supersymmetry in superspace.  However, confirmation of this 
remains for the future.   The problem of classifying 11D superspace 
irreducible representations remains  an important unsolved puzzle.

We again point out for skeptics that non-minimal 4D, $N$ = 1 superspace
geometry is the avatar to set the pattern to be followed by 11D
supergravity/M-theory superspace.  It is a demonstrable fact that imposing
the {\em {solely}} conventional constraints leads to four field strength
superfields; ${\cal W}{}_{a  b c d}$, ${\cal J}_{\a}$,
$X_{\[a b\]}{}^c$ and $X_{\[a b d e f\]}{}^c$, one on-shell and three
off-shell field strengths.  The first of these  contains all the usual
conformal degrees of freedom, the second allows  for the traditional
realization of the Weyl symmetry within a Poincar\' e superspace and the
latter two superfield are roughly speaking the analogs of $G_{\un a}$ in
the non-minimal 4D, $N$  = 1 superspace supergravity theory.  However,
$X_{\[a b\]}{}^c$ and $X_{\[a b d e f\]}{}^c$ differ  from $G_{\un a}$ in
the important respect that they are also conformally  covariant like ${\cal
W}{}_{a b c d}$. The superfield ${\cal W}{}_{a b c d}$ has conformal weight
one  while $X_{\[a b\]}{}^c$ and $X_{\[a b d e f\]}{}^c$ have conformal
weight zero.  In terms of semi-prepotentials, all the dynamics are
contained in two superfields $H_{\a}{}^c$ and $\Psi$. The first of these 
is the gauge field for the Weyl degrees of freedom and the second is the
Goldstone superfield for breaking superconformal symmetry to super
Poincar\' e symmetry. 

Some years ago, we also conjectured \cite{KAPZ} that the ultimate
formulation of covariant string field theory must be one in which the
fundamental string field functional should appear as a {\em
{pre-potential}} in an  as-yet undiscovered geometrical formulation.  In
the most complete formulation of open string field theory to date
\cite{BRK}, Berkovits has presented a formulation that is hauntingly
reminiscent of the 4D, N = 1 pre-potential formulation of Yang-Mills
theory.  This is  precisely in accord with our conjecture in \cite{KAPZ}
and encourages our belief in a type of universality of pre-potential
formulations in all theories that contain supergravity.
 
It has been the suggestion of Cederwall, Gran, Nielsen and Nilsson
\cite{CGNN} that the quantities $X_{\[a b\]}{}^c$ and $X_{\[a b d e f
\]}{}^c$ are both equations of motion.  In the work of \cite{GN01} we 
have pointed out that if one can be set equal to zero as a
constraint that the choice $X_{\[a b d e f
\]}{}^c  = 0$ leads to a smaller supergravity multiplet (if this
is a viable option).  It also the case that both of these $X$-field
strength superfields {\em {cannot}} simultaneously be set to zero as
constraints.  To do so would  force the conformal semi-prepotential $H_{\a
}{}^c$ to be zero up to a pure gauge transformation.

If 11D supergravity is like (i.\ e.\ separable) other prepotential 
formulation of supergravity theories that possess both a conformal 
multiplet and a conformal compensator then the latter will necessarily
demand non-vanishing torsion at dimension 1/2.  The 11D T--C${}^2$ 
theorem is the formal statement of the most fundamental reason why 
it has been our position that Howe's Theorem is specious.  Stated
another way, if Howe's Theorem were true and the $X$-tensors were
the only ones required to completely describe the dynamics of the
M-theory 11D effective action, then this effective action would 
possess a superconformal symmetry.  Not possessing a clear understanding 
of this has led to a number of confused efforts in the research literature
by  different research groups working on both 11D and 10D superspace
problems \cite{BPTFFP}.

In this present work we have found evidence that in 11D supergravity, 
the canonical split of  the fundamental degrees of freedom into a
superconformal pre-potential $H_{\a}{}^c$ and a conformal compensator
$\Psi$ is valid and that it is possible to conventionally realize a
superconformal symmetry.  As 11D supergravity has been proposed as a 
limit of M-theory, our findings naturally suggest that this canonical 
split likely carries over to the complete M-theory itself!  That is, 
it appear likely to us that M-theory has a similar split and that it 
is possible to realize a superconformal symmetry on whatever may play 
the role of the fundamental degrees of freedom that describe M-theory. 
Thus, the concept of the pre-potential will likely play an important 
role in the final formulation of M-theory.  We thus conjecture that 
there exist some formulation of M-theory that possesses all the 
canonical structures of superspace supergravity; a conformal pre-potential,
a conformal compensating pre-potential and a realization of 
superconformal symmetry.

${~~~}$ \newline
${~~~~~~~~~}$``{\it {Understanding and wisdom only comes to us when our 
respective muses \newline ${~~~~~~~\,~~}$ deign to receive an 
audience. Knowing this is the beginning of both.}}'' \newline
${~~~~~~~~~~~}$ -- Anonymous

${~~~}$\newline 
\noindent
{\bf {Acknowledgment}} \newline \noindent
${~~~~}$We wish to acknowledge discussions with M.\ Cederwall and 
P.\ Howe, B.\ Nilsson and H.\ Nishino.

\noindent
{\bf {Added Note in Proof}} \newline \noindent
${~~~~}$After the completion of this work, we recieved communications
from B.\ Nilsson stating that the issue of superconformal symmetry
leads one to  ``... reconsider Weyl superspace as soon as one
puts in explicit expressions for the $X$-tensors.'' As well their
work requires a compution of the Weyl-connection to verify the
validity of `Weyl superspace'.

A component level discussion of 11D superconformal symmetry
may be found in our final reference.

\noindent{{\bf {Appendix : Conventions and Notation}}}

~~~~The conventions that we use for 11D superspace have been
stated in some detail in our previous work \cite{G6}.  In particular, 
we use real (i.e. Majorana) 32-component spinors for the Grassmann
coordinates of superspace.  Our $\g$-matrices are defined by
$$
\{ ~ \g^a \, \, , \, \g^b ~ \} ~=~ 2 \, \eta^{a b} \, {\rm I}
~~~,  \label{eq:AA1}  \eqno(A.1) $$
where the signature of the metric is the ``mostly minus one,''
i.\ e.\ diag. $(+, \, -,\, \cdots ,\, -)$.  This implies that our
gamma matrices satisfy the complex conjugation conditions
$$ \eqalign{
[\, (\g^a )_{\a}{}^{\b} \,]^* &=~ -\, (\g^a )_{\a}{}^{\b} ~~~~,~~~
[\, (\g^{[2]} )_{\a}{}^{\b} \,]^* ~=~  (\g^{[2]} )_{\a}{}^{\b} ~~~,
\cr
[\, (\g^{[3]} )_{\a}{}^{\b} \,]^* &=~ -\, (\g^{[3]} )_{\a}{}^{\b} 
~~~,~~~
[\, (\g^{[4]} )_{\a}{}^{\b} \,]^* ~=~  (\g^{[4]} )_{\a}{}^{\b} ~~~,
\cr
[\, (\g^{[5]} )_{\a}{}^{\b} \,]^* &=~ -\, (\g^{[5]} )_{\a}{}^{\b} 
~~~.
} \label{eq:AA2}  \eqno(A.2) $$
Our `spinor metric,' with which we raise and lower spinor indices,
is denoted by $C_{\a \b}$ and satisfies,
$$
C_{\a \b} ~=~ - \, C_{\b \a} ~~~,~~~ [\, C_{\a \b} \,]^* ~=~ - 
\, C_{\a \b} ~~~.
\label{eq:AA3}   \eqno(A.3) $$
The inverse spinor metric $ C^{\a \b} $ is defined to satisfy
$$
C_{\a \b} \, C^{\g \b} ~=~ \d_{\a}{}^{\g} ~~~. 
\label{eq:AA4}  \eqno(A.4) $$
We also use superspace conjugation which permits the appearance of 
appropriate factors of $i$ even within a theory of involving solely
real spinors.  A complete discussion of superspace conjugation can
be found  in a recent pedagogical presentation \cite{TASI} (p. 13).

We define our gamma matrices with multiple numbers of vector
indices through the equations
$$ \eqalign{
\g_a \, \g_b &=~ \g_{a \, b} ~+~ \eta_{ a \,b} ~~~,~~~~~~~~~~~~~~
~~~~~~~ \g_b \, \g_a ~=~ -\, \g_{a \, b} ~+~ \eta_{ a \,b} ~~~, \cr
\g_a \, \g_{b \, c} &=~ \g_{a \, b \, c} ~+~ \eta_{ a \, [ b}  
\g_{c]} ~~~, ~~~~~~~~~~~~\,~~ \g_{b \, c}\, \g_a ~=~  \g_{a \, b 
\, c} ~-~ \eta_{ a \,  [ b}  \g_{c]} ~~~, \cr
\g_a \, \g_{b \, c \, d} &=~ \g_{a \, b \, c \, d} ~+~ \fracm 12 
\eta_{ a \, [ b|} \g_{| c \, d] } ~~~, ~~~~~~~\g_{b \, c \, d} \, 
\g_a ~=~ - \,  \g_{a \, b \, c \, d} ~+~ \fracm 12 \eta_{ a \, [ b|}
\g_{| c \, d] } ~~~,  \cr
{~~} \g_a \, \g_{b \, c \, d \, e} &=~ \g_{a \, b \, c \, d \, e} 
~+~ \fracm 16 \eta_{ a \, [ b|} \g_{| c \, d \, e] }
~~~, ~~~ \g_{b \, c \, d \, e}\, \g_a ~=~ \g_{a \, b \, c \, d \, 
e} ~-~ \fracm 16 \eta_{ a \, [ b|} \g_{| c \, d \, e] }
~~~, \cr
&{~~} \g_a \, \g_{b \, c \, d \, e \, f} =~  i \, \fracm 1{120} \,
\e_{a \, b \, c \, d \, e \, f}{}^{[5]} \g_{[5]} ~+~ \fracm 1{24} 
\eta_{a \,  [ b|} \g_{| c \, d \, e \, f] }  ~~~, \cr
&{~~} \g_{b \, c \, d \, e \, f} \, \g_a  =~ - i \, \fracm 1{120} 
\, \e_{a \, b \, c \, d \, e \, f}{}^{[5]} \g_{[5]} ~+~ \fracm 1{24}
\eta_{ a \,  [ b|} \g_{| c \, d 
\, e \, f] }  ~~~. 
} \label{eq:AA5}  \eqno(A.5) $$ 

The basic identities for non-vanishing traces over the gamma matrices 
are
$$  \eqalign{
{~~~~~~}&\fracm 1{32} \, {\rm {Tr}}\,(\, \g_a \, \g^b \,) ~=~ \delta_{
a}{}^{b} ~~~, \cr 
&\fracm 1{32} \, {\rm {Tr}}\,(\, \g_{a_1 a_2} \, \g^{b_1 b_2} \,)~=~  
- \, \d_{[ a_1|}{}^{b_1} \, \d_{| a_2]}{}^{b_2}  ~~~, \cr 
&\fracm 1{32} \, {\rm {Tr}}\,(\, \g_{a_1 a_2 a_3} \, \g^{b_1 b_2 b_3} 
\,) ~=~ - \, \d_{[ a_1|}{}^{b_1} \, \d_{| a_2|}{}^{b_2}  \,  \d_{|
a_3]}{}^{b_3}    ~~~,\cr 
&\fracm 1{32} \, {\rm {Tr}}\,(\, \g_{a_1 a_2 a_3 a_4} \, \g^{b_1 b_2 
b_3 b_4} \,) ~=~  \d_{[ a_1|}{}^{b_1} \, \d_{| a_2|}{}^{b_2}  
\, \d_{| a_3|}{}^{b_3} \, \d_{| a_4]}{}^{b_4}~~~,   \cr
&\fracm 1{32} \, {\rm {Tr}}\,(\, \g_{a_1 a_2 a_3 a_4 a_5} \, \g^{b_1 
b_2 b_3 b_4 b_5} \,) ~=~  \d_{[ a_1|}{}^{b_1} \, \d_{| a_2|}{
}^{b_2} \, \d_{| a_3|}{}^{b_3} \, \d_{| a_4|}{}^{b_4}  \, \d_{| a_5]
}{}^{b_5} ~~~,\cr 
&\fracm 1{32} \, {\rm {Tr}}\,(\, \g_{a_1} \,\cdots \, \g_{a_{11}} )
~=~  i \,  \e_{a_1 \dots a_{11}}  ~~~.\cr
} \label{eq:AA6a}  \eqno(A.6) $$

Using the spinor metric to lower one spinor index of the quantities 
in (A.2) we find
$$ \eqalign{
[\, (\g^a )_{\a \b} \,]^* &=~  (\g^a )_{\a\b} ~~~~,~~~
[\, (\g^{[2]} )_{\a \b} \,]^* ~=~  - \, (\g^{[2]} )_{\a \b} ~~~,
\cr
[\, (\g^{[3]} )_{\a \b} \,]^* &=~  (\g^{[3]} )_{\a\b} ~~~,~~~
[\, (\g^{[4]} )_{\a \b} \,]^* ~=~ - \, (\g^{[4]} )_{\a \b} ~~~,
\cr
[\, (\g^{[5]} )_{\a \b} \,]^* &=~ (\g^{[5]} )_{\b \a} ~~~,
} \label{eq:AA7}  \eqno(A.7) $$
and as well the same equations apply to the matrices with two raised 
spinor indices.  In addition these satisfy the symmetry relations
$$ \eqalign{ {~~~~~~~}
 (\g^a )_{\a \b}  &=~  (\g^a )_{\b\a} ~~~~,~~~
(\g^{[2]} )_{\a \b}  ~=~   (\g^{[2]} )_{\b \a} ~~~,~~~
(\g^{[5]} )_{\a \b}  ~=~ (\g^{[5]} )_{\b \a} ~~~,
\cr
(\g^{[3]} )_{\a \b} &=~  - \, (\g^{[3]} )_{\b\a} ~~~,~~~
(\g^{[4]} )_{\a \b} ~=~ - \, (\g^{[4]} )_{\b \a} ~~~,
} \label{eq:AA8}  \eqno(A.8) $$
where the same equations apply to the matrices with two raised spinor 
indices.

Other identities on the 11D gamma matrices include
$$
\g^{[P]} \, \g_{[Q]} \, \g_{[P]} ~=~ c_{[Q] \, [P]} \, \g_{[Q]} ~~~,
\label{eq:AA9}  \eqno(A.9) 
$$
where the coefficients $c_{[Q] \, [P]}$ are given in the following
table.
\vspace{0.1cm}
\begin{center}
\footnotesize
\begin{tabular}{|c|c|c|c|c|c|}\hline
$~$ & $[P] = [1]$ & $[P] = [2]$ & $[P] = [3]$ & $[P] = [4]$ & 
$[P] = [5]$  \\ \hline
$ [Q] = [1] $ & $ - 9 $ & $ - 70 $ & $ 450 $ & $ 2,160 $ & $ - 5,040 $  
\\ \hline
$ [Q] = [2] $ & $ 7 $ & $ - 38 $ & $ - 126 $ & $ - 144 $ & $ - 5,040 $  
\\ \hline
$ [Q] = [3] $ & $ - 5 $ & $ - 14 $ & $ - 30 $ & $ - 528 $ & $ 1,680 $  
\\ \hline
$ [Q] = [4] $ & $ 3 $ & $ 2 $ & $ 66 $ & $ - 144 $ & $ 1,680 $  
\\ \hline
$ [Q] = [5] $ & $ - 1 $ & $ 10 $ & $ - 30 $ & $ 240 $ & $ - 1,200 $  
\\ \hline
\end{tabular}
\end{center}
For example this table implies
$$
\g^{[3]} \, \g_{[2]} \, \g_{[3]} ~=~ - 126 \, \g_{[2]} ~~~,
\label{eq:AA10}  \eqno(A.10) 
$$

Some useful Fierz-type identities include the following
$$ \eqalign{ {~~~~~~~}
0\, &=~  (\g^a)_{( \a \b |} (\g_{a b})_{| \g \d )}   ~~~, \cr
0\, &=~ 5 (\g^a)_{( \a \b |} (\g_a)_{| \g \d )} ~+~ \fracm 12 (\g^{
[2]})_{( \a \b |} (\g_{[2]})_{| \g  \d )} ~~~, \cr
0\, &=~ 6 (\g^a)_{( \a \b |} (\g_a)_{| \g \d )} ~+~ \fracm 1{5!} 
(\g^{[5]})_{( \a \b |} (\g_{[5]})_{| \g  \d )} ~~~, \cr
0\, &=~  (\g^e)_{( \a \b |} (\g_{a b c d e})_{| \g \d )} ~-~ \fracm 
1{8} (\g_{[ a b |})_{( \a \b |} (\g_{| c d]})_{| \g  \d )} ~~~, \cr
0\, &=~ \fracm 12 (\g^{a b})_{( \a \b |} (\g_{a b c d e})_{| \g 
\d )} ~+~ 2 (\g_{[ c |})_{( \a \b |} (\g_{| d e ]})_{| \g  \d )} 
~~~, \cr
0\, &=~ (\g^a)_{( \a \b |} (\g_a)_{| \g )}{}^{\d} ~-~ \fracm 12 
(\g^{[2]})_{( \a \b |} (\g_{[2]})_{| \g  )}{}^{\d} ~+~ \fracm 
1{5!} (\g^{[5]})_{( \a \b |} (\g_{[5]})_{| \g  )}{}^{\d} ~~~.
 }\label{eq:AA11}  \eqno(A.11) $$

Our Lorentz generator is defined to realize
$$ \eqalign{ {~~~~}
[\, {\cal M}{}_{a b} ~&,~ (\g_{a})_{\a \b} \, \} ~=~ 0 ~~~, \cr
[\, {\cal M}{}_{a b} ~&,~ \nabla_{\a} \, \} ~=~ \fracm 12 \,
(\g_{a b})_{\a}{}^{\b} \, \nabla_{\b} ~~~, ~~~
[\, {\cal M}{}_{a b} ~,~ \nabla_{c} \, \} ~=~ \eta_{c a} \, 
\nabla_b ~-~ \eta_{c b} \, \nabla_a  ~~~, \cr 
[\, {\cal M}{}_{a b} ~&,~ {\cal M}{}_{c d} \, \} ~=~ \eta_{c a} \, 
{\cal M}{}_{b d} ~-~ \eta_{c b} \, {\cal M}{}_{a d} ~-~ \eta_{d a} 
\, {\cal M}{}_{b c} ~+~ \eta_{d b} \, {\cal M}{}_{a c} ~~~.
}\label{eq:AA12}  \eqno(A.12)
$$

One notational device not discussed in our previous paper is 
the definition of the 11D super-epsilon tensor, which we denote 
by ${\Hat \e}{}_{A_1 \dots A_4}{}^{B_1 \dots B_7} $, and define 
by
$$ \eqalign{
{\Hat \e}{}_{a_1 \, \dots a_{4}}{}^{b_1 \, \dots b_{7}} &\equiv~ 
\e_{a_1 \, \dots a_{4}}{}^{b_1 \, \dots b_{7} } ~~~,  \cr
{\Hat \e}{}_{\a_1 \, a_2 \dots a_{4}}{}^{ \b_1 \, b_2 \dots b_{7}} 
&\equiv~ (\g^{a_1}{}_{b_1})_{\a_1}{}^{\b_1}\, \e_{a_1 \, \dots 
a_{4}}{}^{b_1 \, \dots b_{7} }  ~~~,  \cr
{\Hat \e}{}_{\a_1 \, \a_2 \, a_3 \, a_{4}}{}^{\b_1 \, \b_2 \, b_3
 \dots b_{7}}  &\equiv\, (\g^{a_1}{}_{b_1})_{(\a_1|}{}^{\b_1} 
\,(\g^{a_2}{}_{b_2})_{|\a_2)}{ }^{\b_2}\, \e_{a_1 \, \dots a_{4}}
{}^{ b_1 \, \dots b_{7} } ~~~, \cr 
{\Hat \e}{}_{\a_1 \, \a_2 \, \a_3 \, a_{4}}{}^{\b_1 \, \b_2 \, 
\b_3 \, b_4 \, b_5 \, b_6 \, b_{7} } &\equiv\, (\g^{a_1}{}_{b_1
})_{(\a_1|}{}^{\b_1}\, (\g^{a_2}{}_{b_2})_{|\a_2|}{}^{\b_2}\, 
(\g^{a_3}{}_{b_3})_{|\a_3)}{}^{\b_3}\, \e_{a_1 \, \dots a_{4}}
{}^{ b_1 \, \dots b_{7} } ~~~,  \cr 
{\Hat \e}{}_{\a_1 \, \a_2 \, \a_3 \, \a_{4}}{}^{\b_1 \, \b_2 \, 
\b_3 \, \b_4 \, b_5 \, b_6 \, b_{7}} &\equiv\, (\g^{a_1}{}_{b_1
})_{(\a_1|}{}^{\b_1}\, (\g^{a_2}{}_{ b_2})_{|\a_2 |}{}^{\b_2}\,
(\g^{a_3}{}_{b_3})_{|\a_3 |}{}^{\b_3} \, (\g^{a_4}{}_{b_4})_{|
\a_4)}{}^{\b_4}\, \e_{a_1 \, \dots a_{4}}{}^{b_1 \, \dots b_{7} 
} ~~~. } 
\label{eq:AA13} \eqno(A.13) $$  
The role of this object is that it allows us to convert the components
of a super 7-form $X_{A_1 \dots A_7}$ into the dual components of 
a super 4-form ${\Hat X}{}_{A_1 \dots A_4}$ via the definitions
$$ \eqalign{
{\Hat X}{}_{a_1 \dots a_4} &\equiv~ \fracm 1{7!} {\Hat \e}{}_{
a_1 \dots a_4}{}^{b_1 \dots b_7} X_{b_1 \dots b_7} ~~~, \cr
{\Hat X}{}_{\a_1 a_2 a_3 a_4} &\equiv~ \fracm 1{ 6!} {\Hat \e}{
}_{\a_1 a_2 a_3 a_4}{}^{\b_1 b_2 \dots b_7}  X_{\b_1 b_2 \dots 
b_7} ~~~~, \cr 
{\Hat X}{}_{\a_1 \a_2 a_3 a_4} &\equiv~ \fracm 1{2 \cdot 5!}   
{\Hat \e}{}_{\a_1 \a_2 a_3 a_4}{}^{\b_1 \b_2 b_3 \dots b_7} X_{
\b_1 \b_2 b_3 \dots b_7}  ~~~~, \cr
{\Hat X}{}_{\a_1 \a_2 \a_3 a_4} &\equiv~ \fracm 1{3! \cdot 4!}   
{\Hat \e}{}_{\a_1 \a_2 \a_3 a_4}{}^{\b_1 \b_2 \b_3 b_4 \dots b_7
} X_{\b_1 \b_2 \b_3 b_4 \dots b_7}  ~~~~, \cr
{\Hat X}{}_{\a_1 \a_2 \a_3 \a_4} &\equiv~ \fracm 1{4! \cdot 3!}   
{\Hat \e}{}_{\a_1 \a_2 \a_3 \a_4}{}^{\b_1 \b_2 \b_3 \b_4 b_5 b_7 
b_7}  X_{\b_1 \b_2 \b_3 \b_4 b_5 b_6 b_7} ~~~. \cr
} \label{eq:AA14}  \eqno(A.14) $$
As well, it can be used for the reverse purpose of converting the 
components of a super 4-form to those of a super 7-form.  The
super-epsilon tensor concept has proven very useful for 10D theories 
as we suspect will also be  the case for 11D theories.



\begin{thebibliography}{66}

\bibitem{GN01}S.\ J.\ Gates, Jr. and H.\ Nishino, ``Deliberations 
on 11D Superspace for the M-Theory Effective Action,'' Univ. of MD 
preprint UMDEPP 00-032, hep-th/0101037 (revised).

\bibitem{G1}S.\ J.\ Gates, Jr. and H.\ Nishino, Phys.\ Lett.\ 
{\bf {173B}} (1986) 52.

\bibitem{G2}S.\ J.\ Gates, Jr. and S.\ I.\ Vashakidze, Nucl.\ 
Phys.\ {\bf {B291}} (1987) 173;S.\ J.\ Gates, Jr. and H.\ Nishino, 
Nucl.\ Phys.\ {\bf {B291}} (1987) 205. 

\bibitem{G3}S.\ Bellucci and S.\ J.\ Gates, Jr., Phys.\ Lett.\ 
{\bf {208B}} (1988) 456. 

\bibitem{G4} S.\ Bellucci, S.\ J.\ Gates, Jr.\ and D.\ Depireux, 
Phys.\ Lett.\ {\bf {238B}} (1990) 315.

\bibitem{G5}S.\ Bellucci, S.\ J.\ Gates, Jr.\ , B.\ Radak, P.\ 
Majumdar and S.\ Vashakidze, Mod.\ Phys.\ Lett.\ {\bf {A21}} (1989) 
1985.

\bibitem{G80}S.\ J.\ Gates, Jr., Phys.\ Lett.\ {\bf {96B}} (1980) 
305.

\bibitem{G6}H.\ Nishino and S.\ J.\ Gates, Jr., Phys.\ Lett.\ 
{\bf {388B}} (1996) 504.

\bibitem{HOW}P.\ Howe, Phys.\ Lett.\ {\bf {415B}} (1997) 149 
(hep-th/9707184). 

\bibitem{GG}S.\ J.\ Gates, Jr. and R.\ Grimm, Phys.\ Lett.\ {\bf {133B}}
(1983) 192.

\bibitem{ChrLcomp}W.\ Siegel, ``A Derivation of the Supercurrent
Superfield,'' HUTP-77/A089, Dec 1977; ibid.\ ``The Superfield
Supergravity Action,'' HUTP-77/A080, Dec 1977; idem.\
Nucl.\ Phys.\ {\bf {B142}} (1978) 301.

\bibitem{HOW1}P.\ Howe and R.\ Tucker, Phys.\ Lett.\ {\bf {80B}}
(1978) 138.
 
\bibitem{barHOW}W.\ Siegel, Phys.\ Lett.\ {\bf {80B}} (1979) 224; 
S.\ J.\ Gates, Jr., in {\it Supergravity} eds.\ P.\ van 
Nieuwenhuizen and D.\ Z.\ Freedman, North-Holland, Amsterdam,
1979, pp. 215-219; idem.\ Nucl.\ Phys.\ {\bf {B162}} (1980) 79; 
ibid.\ Nucl.\ Phys.\ {\bf {B176}} (1980) 397; P.\ Howe, Phys.\ 
Lett.\ {\bf {100B}} (1980) 389; idem.\ , Nucl.\ Phys.\ {\bf 
{199}} (1982) 309.

\bibitem{Gconf}S.\ J.\ Gates, Jr.\ and H.\ Nishino, Phys.\ Lett.\ 
{\bf {266B}} (1991) 14.

\bibitem{CGNN}M.\ Cederwall, U.\ Gran, M.\ Nielsen and B.\ Nilsson,
``Manifestly Supersymmetric M-Theory,'' G\" oteborg ITP preprint,
hep-th/007035.

\bibitem{GN85}S.\ J.\ Gates, Jr.\ and H.\ Nishino, Class. and Quant.\ 
Grav.\ {\bf {3}} (1986) 391.

\bibitem{ConSTRT}M.\ Brown and S.\ J.\ Gates, Jr., Annals of Physics, 
{\bf {122}}, No. 2 (1979) 443; S.\ J.\ Gates, Jr.\ and W.\ Siegel, 
Nucl.\ Phys.\ {\bf {B163}} (1980) 519; S.\ J.\ Gates, Jr., M.\ T.\ 
Grisaru, M.\ Ro\v cek and W.\ Siegel, ``Superspace or One Thousand 
and One Lessons in Supersymmetry," Benjamin Cummings (Addison-Wesley), 
Reading, MA (1983).

\bibitem{GSW}P. Breitenlohner, Phys.\ Lett.\ {\bf {67B}} (1977) 49; idem.\
Nucl.\ Phys.\ {\bf {B124}} (1977) 500; W.\ Siegel, Harvard preprint {\bf
{HUTP-77/A068}} (November, 1977) and Harvard preprint {\bf {HUTP-77/A080}}
(November, 1977); S.\ J.\ Gates, Jr.\ and J.\ A.\ Shapiro, Phys.\ Rev.\ {\bf 
{D18}} (1978) 2768; W.\ Siegel and S.\ J.\ Gates, Jr., Nucl.\ Phys.\ {\bf 
{B147}} (1979) 77; S.\ J.\ Gates, Jr.\ and M.\ Brown, Nucl.\ Phys.\ {\bf 
{B165}} (1980)  445.

\bibitem{PVW}K.\ Peeters, P.\ Vanhove and A.\ Westerberg, 
``Supersymmetric $R^4$ Actions and Quantum corrections to 
Superspace Torsion Constraints,'' hep-th/0010182; idem. 
``Supersymmetric Higher-derivative Actions in Ten and Eleven 
Dimensions, the Associated Superalgebras and Their Formulation 
in Superspace', hep-th/0010167. 

\bibitem{GR}S.\ J.\ Gates, Jr.\ and L.\ Rana, ``A Primer on 
Supersymmetric Quantum Mechanics (I),'' preprint UMDEPP 96-38,
in preparation.

\bibitem{TASI}S.\ J.\ Gates, Jr., ``Basic Canon in D = 4, $N$ = 1
Superfield Theory: Five Primer Lectures,'' in {\em {Supersymmetry,
Supergravit and Supercolliders: TASI 97}}, ed.\ J.\ Bagger, World
Scientific, Singapore, 1999, hep-th/9809064 .

\bibitem{2DN4}S.\ J.\ Gates, Jr., L.\ Lu and R.\ Oerter, Phys.\
 Lett.\ {\bf {218B}} (1989) 33.

\bibitem{KAPZ}S.\ J.\ Gates, Jr., ``Strings, Superstrings and 
Two-Dimensional Lagrangian Field  Theory'', in {\underline {Functional}} 
{\underline {Integration}}, {\underline {Geometry}} {\underline  {and}}
{\underline {Strings}}, the {\it Proceedings of the XXV  Winter School of
Theoretical Physics} in Karpacz, Poland (Feb.,  1989) Z.Haba and J.Sobczyk,
Birkhauser-Verlag Press (1989) pp.  140-184.

\bibitem{BRK}N.\ Berkovits, ``Review of Open Superstring Field
Theory,'' (hep-th/0105230).

\bibitem{BPTFFP}L.\ Bonora, P.\ Pasti, and M.\ Tonin, Phys.\ Lett.\ 
{\bf {188B}} (1987) 335; S.\ Ferrara, P.\ Fr\' e and M.\ Porrati, Ann.\ 
Phys.\ {\bf {175}} (1987)  112.

\bibitem{BRGDRU}E.\ Bergshoeff and M.\ de Roo, Phys.\ 
Lett.\ {\bf {138B}} (1984) 67.


\end{thebibliography}
\end{document}